\documentclass[a4paper, amsfonts, amssymb, amsmath, reprint, showkeys, nofootinbib, twoside, superscriptaddress, footnotes]{revtex4-1}

\usepackage[T1]{fontenc}
\usepackage{float}
\usepackage[version=4]{mhchem}
\usepackage{caption}
\usepackage{subcaption}
\usepackage{nicematrix}
\usepackage{physics}
\usepackage{float}

\usepackage{algorithm}
\usepackage{algpseudocode}

\usepackage[unicode=true, colorlinks=true]{hyperref} 

\usepackage{booktabs} 

\usepackage{tikz}
\usetikzlibrary{shapes.geometric, arrows}
\tikzstyle{startstop} = [rectangle, rounded corners, minimum width=3cm, minimum height=1cm,text centered, draw=black, fill=red!30]
\tikzstyle{io} = [trapezium, trapezium left angle=70, trapezium right angle=110, minimum width=3cm, minimum height=1cm, text centered, draw=black, fill=blue!30]
\tikzstyle{process} = [rectangle, minimum width=3cm, minimum height=1cm, text centered, text width=3cm, draw=black, fill=orange!30]
\tikzstyle{decision} = [diamond, minimum width=2cm, minimum height=0.5cm, text centered, text width=2cm, draw=black, fill=green!30, aspect=1.75, inner sep=-0.5ex]
\tikzstyle{arrow} = [thick,->,>=stealth]
\tikzstyle{dotted-arrow} = [arrow, dotted]

\begin{document}
\title{Bridging Quantum Chemistry and MaxCut: Classical Performance Guarantees and Quantum Algorithms for the Hartree-Fock Method}

\author{Alexis Ralli}
\affiliation{Department of Physics and Astronomy, Tufts University, Medford, MA 02155, USA}
\affiliation{Centre for Computational Science, Department of Chemistry, University College London, WC1H 0AJ, United Kingdom}
\author{Tim Weaving}
\affiliation{Centre for Computational Science, Department of Chemistry, University College London, WC1H 0AJ, United Kingdom}
\author{Peter V. Coveney}
\affiliation{Centre for Computational Science, Department of Chemistry, University College London, WC1H 0AJ, United Kingdom}
\affiliation{Advanced Research Computing Centre, University College London, WC1H 0AJ, United Kingdom}
\affiliation{Informatics Institute, University of Amsterdam, Amsterdam, 1098 XH, Netherlands}
\author{Peter J. Love}
\affiliation{Department of Physics and Astronomy, Tufts University, Medford, MA 02155, USA}
\affiliation{Computational Science Initiative, Brookhaven National Laboratory, Upton, NY 11973, USA}

\date{\today}

\begin{abstract}
In quantum chemistry, self-consistent field (SCF) algorithms define a nonlinear optimization problem, with both continuous and discrete components. In this work, we derive Hartree-Fock-inspired SCF algorithms that can be exactly written as a sequence of Quadratic Unconstrained Spin/Binary Optimization problems (QUSO/QUBO). We reformulate the optimization problem as a series of MaxCut graph problems, which can be efficiently solved using semi-definite programming techniques. This procedure provides performance guarantees at each SCF step, irrespective of the complexity of the optimization landscape. We numerically demonstrate the QUBO-SCF and MaxCut-SCF methods by studying the hydroxide anion \ce{OH-} and molecular Nitrogen \ce{N2}. The largest problem addressed in this study involves a system comprised of 220 qubits (equivalently, spin-orbitals). Our results show that QUBO-SCF and MaxCut-SCF suffer much less from internal instabilities compared with conventional SCF calculations. Additionally, we show that the new SCF algorithms can enhance single-reference methods, such as configuration interaction.  Finally, we explore how quantum algorithms for optimization can be applied to the QUSO problems arising from the Hartree-Fock method. Four distinct hybrid-quantum classical approaches are introduced: GAS-SCF, QAOA-SCF, QA-SCF and DQI-SCF.








\end{abstract}

\maketitle

\section{Introduction} \label{sec:introduction}


Over the past decade, quantum computing has achieved remarkable advancements, with Noisy Intermediate-Scale Quantum (NISQ) systems featuring hundreds of qubits and quantum annealing systems comprising thousands of qubits. One of the major potential applications of quantum computing is quantum simulation, particularly for quantum chemistry \cite{cao2019quantum} and optimization problems \cite{abbas2024challenges}. Quantum algorithms for optimization problems include the Quantum Approximate Optimization Algorithm (QAOA) \cite{farhi2014quantum}, Quantum Annealing (QA) \cite{apolloni1989quantum, apolloni1990numerical}, Decoded Quantum Interferometry (DQI) \cite{jordan2024optimization}, Grover Search \cite{grover1996fast} and Quantum Adiabatic Optimization (QAO) \cite{reichardt2004quantum}.

Both QUBO and QUSO problems are a class of combinatorial optimization problems. In general, these problems are NP-hard \cite{boros2020compact}; however, many heuristics have been developed to approximately solve them. Examples include: Simulated Annealing \cite{alkhamis1998simulated}, Branch-and-Cut methods \cite{barahona1989experiments, pardalos1990computational}, Semi-Definite Programming \cite{helmberg1998solving} and many more. There have also been quantum-inspired classical heuristics, such as the Simulated Bifurcation Algorithm (SBA) \cite{goto2019combinatorial}.

An alternative way to solve such optimization problems is via a purpose-built computer. The high level idea is to map the second degree pseudo-Boolean function onto an Ising Hamiltonian and allow that system evolve into the ground state, thus solving the problem. D-Wave has built superconducting flux qubit quantum annealers, which can be used to solve such Ising problems \cite{kadowaki1998quantum}. Fujitsu have built a ``Digital Annealing Unit" (DAU) \cite{matsubara2020digital}. Benchmark results indicate that Fujitsu's new computer architecture performed on par with, or better than, the general-purpose CPLEX solver and leading specialized solvers. Hitachi has produced a Complementary Metal Oxide Semiconductor (CMOS) annealing device that artificially reproduces the behavior of the Ising model along a CMOS circuit \cite{yamaoka201524}. Nippon Telegraph and Telephone (NTT) Corporation, in collaboration with the National Institute of Informatics (NII), built a coherent Ising machine using a network of $100,512$ degenerate optical parametric oscillator pulses working as the Ising spins~\cite{honjo2021100}. 

One issue with using specialist Ising solver hardware is a limitation on the connectivity between spin sites. For example D-Wave's 5,000-qubit device is restricted to $15$ connections per qubit. This results in two problems: (1) A theoretical problem, whereby a given Ising system needs to be mapped onto a subgraph of the physical hardware; (2) An  engineering problem, where increased connectivity between spin-sites will allow more complex graphs to be described. Improvements in either/or both areas will allow larger systems to be studied.


Most quantum algorithms for chemistry focus on the high-accuracy end of the spectrum for orbital-based methods. Full Configuration Interaction (FCI) has been a particular target of quantum algorithm development since the first paper on the topic~\cite{aspuru2005simulated}. Self-consistent field algorithms such as Hartree-Fock (HF) are of much lower accuracy. However, it is important to note that any improvements to HF problems will have an impact on all quantum chemistry methods that use a basis of HF orbitals as a starting point. This includes all orbital-based post-Hartree-Fock wavefunction methods such as DMRG, Coupled Cluster, Configuration Interaction and M{\o}ller-Plesset perturbation theory.

Hartree-Fock theory is the foundation of molecular orbital (MO) theory \cite{hartree1928wave, fock1930naherungsmethode, roothaan1951new, hall1951molecular}. It provides a method to approximate the electronic wave functions of molecules by describing each electron as occupying a single molecular orbital. The key principle of the Hartree-Fock method is that each electron is governed by a single-particle Schr\"{o}dinger equation (the Fock operator) \cite{Pople1970, szabo2012modern, helgaker2013molecular}. The mathematical problem is written as a set of coupled integro-differential equations. This can be solved numerically, but is usually only possible for simple systems due to the large number of basis functions required by real-space approaches \cite{lehtola2019review}.

Instead, the spin-orbitals can be projected onto a finite set of known basis functions, a Galerkin approximation \cite{Fletcher1984, Fontana2005}, which converts this continuous problem into a discrete problem. The integro-differential problem is converted into a system of linear equations, which can then be solved with matrix techniques. 
This defines a pseudo-eigenvalue problem, and because the Fock operator has functional dependence on its eigenfunctions it is a nonlinear problem. The problem is therefore solved via nonlinear techniques. The most common methodology is an iterative procedure that is repeated until the solution converges, colloquially referred to as achieving ``self-consistency". This is why the technique is known as a self-consistent field (SCF) method.


Several algorithms in quantum chemistry utilize combinatorial optimization as a subroutine. Examples include Hartree-Fock (HF) and Kohn-Sham Density Functional Theory (DFT) both of which are NP-complete~\cite{whitfield2014np, whitfield2014computational}. The fact that HF and DFT are NP-complete is a worst-case statement~\cite{whitfield2014np, whitfield2014computational}. It implies there exist instances of these problems that (unless $\mathrm{P} = \mathrm{NP}$) will be intractable for classical, and possibly quantum, computers.  One can also ask about the approximability of combinatorial optimization problems in chemistry. This is of interest even for classical computation where, as we shall show, one can obtain performance guarantees for Hartree-Fock. Furthermore, it is reasonable to consider whether quantum optimization approaches, such as Quantum Annealing, QAOA, or DQI can offer an advantage for SCF algorithms. Quantum annealing, QAOA and DQI provide approximate solutions to optimization problems. Here, improving the quality of the approximation is the target for quantum advantage. Recently quantum speedup has been achieved for one problem by the DQI algorithm~\cite{jordan2024optimization}, and for one ensemble of instances of MaxCut for QAOA~\cite{farhi2025lower}. Combinatorial optimization problems in chemistry provide new ensembles of instances on which to investigate the performance of these quantum algorithms.

Previous works have used Quantum Annealing to address the electronic structure problem in various ways~\cite{xia2017electronic,copenhaver2021using, chermoshentsev2021polynomial, genin2019quantum}. Xia, Bian and Kais mapped the full electronic structure Hamiltonian to a many-body Ising Hamiltonian and auxiliary qubits were introduced to reduce the locality to two-body interactions~\cite{xia2017electronic}. In the work of Genin, Ryabinkin and Izmaylov an annealer was used to improve a combinatorial optimization step in the coupled-cluster approach~\cite{genin2019quantum}. These two methods have been compared~\cite{copenhaver2021using} and the difficulties of applying annealing directly to the electronic structure problem, arising from the need for auxiliary variables, have been highlighted~\cite{chermoshentsev2021polynomial}.

In the present paper, we combine quantum optimization and quantum chemistry through the optimization of molecular orbitals in quantum chemistry. We focus on these problems as instances of combinatorial optimization problems and obtain performance guarantees for classical algorithms. We numerically investigate a fixed-point electronic structure calculation for the \ce{OH-} anion in the 6-31G basis set ($22$ qubit problem) and the dissociation of molecular \ce{N2} in the \mbox{cc-pVDZ}, \mbox{cc-pVTZ} and cc-pVQZ basis sets ($56$, $120$ and $220$ qubit problems, respectively). We show how single reference SCF quantum chemistry calculations in a finite basis can be exactly written as a Quadratic Unconstrained Spin Optimization (QUSO) problem, which is equivalent to an Ising problem.  No auxiliary variables are required to reduce the locality of the problem as the Hartree-Fock Hamiltonian is already two-local, giving rise to a QUSO directly. However, our methodology is still limited if the output Ising problem cannot be mapped directly onto the hardware due to connectivity constraints between spins. It is possible to remedy this by using ancilla bits to embed the problem onto such machines \cite{okada2019improving}, but this is not desirable as unnecessary degrees of freedom have been added to the problem. Fortunately, if QUBO/Ising solvers are run on conventional computers, no such connectivity issues are present.

In Section \ref{sec:background} we provide all the relevant quantum chemistry background. Section \ref{sec:SCF_model} then introduces the new classical QUSO-,QUBO- and MaxCut-SCF and the hybrid quantum-classical GAS-SCF, QAOA-SCF, QA-SCF and DQI-SCF algorithms.  Section \ref{sec:discussion} outlines the methods used to generate our numerical results. All numerical data were obtained on classical hardware; no quantum computations were performed. We then provide a comparison of our SCF algorithm against conventional Hartree-Fock calculations, giving a discussion of why the new approach is more resistant to internal instabilities. We close the paper with some conclusions and directions for future work in Section~\ref{sec:conclusion}.

\section{Background}\label{sec:background}

In this Section we provide the necessary background for the new methods introduced. We first review Hartree-Fock in Section~\ref{sec:QChem}, and then review quadratic optimization problems on qubits and spins in Section~\ref{sec:QUBO}.

\subsection{Hartree-Fock theory \label{sec:QChem}}

The non-relativistic second-quantized electronic molecular Hamiltonian under the Born-Oppenheimer approximation is written \cite{helgaker2013molecular}:

\begin{equation} \label{eq:spin_free_H}
    \begin{aligned}
        H^{\text{ferm}} &= \sum_{p,q}^{M} h_{pq}(C) \alpha_{p}^{\dagger} \alpha_{q}  + \frac{1}{2} \sum_{p,q,r,s}^{M} g_{pqrs}(C) \alpha_{p}^{\dagger} \alpha_{r}^{\dagger} \alpha_{s} \alpha_{q} \\
        &= \sum_{p,q}^{M} h_{pq} \alpha_{p}^{\dagger} \alpha_{q}  + \frac{1}{2} \sum_{p,q,r,s}^{M} g_{pqrs} \alpha_{p}^{\dagger} \alpha_{r}^{\dagger} \alpha_{s} \alpha_{q}
    \end{aligned}
\end{equation}
Here $\alpha_{i}^{\dagger}$ and $\alpha_{i}$ are fermionic creation and annihilation operators that act on a basis of orthogonal spin-orbitals. $M$ is the number of spin-orbitals, $h_{pq}$ and $g_{pqrs}$ are the one- and two-electron integrals expressed in an orthonormal molecular orbital (MO) basis. These integrals have functional dependence on the molecular orbital coefficient matrix $C$ \cite{szabo2012modern, helgaker2013molecular}; see Appendix \ref{sec: BCH_fermionic}, Equation~\eqref{eq:integrals1_2}. Note that since $C$ (MO basis) is usually fixed for post-Hartree-Fock methods, in most literature the Hamiltonian is written as the second line of Equation~\eqref{eq:spin_free_H}, where the dependence on $C$ is omitted. 

For a fixed $C$, we can obtain a new basis of molecular orbitals by the following unitary rotation \cite{linderberg1977state, dalgaard1978optimization, couty1997generalized, helgaker2013molecular, kossoski2021excited}: 

\begin{equation} \label{eq:U}
    \begin{aligned}
        U(\vec{\kappa}) = e^{-\hat{k}(\vec{\kappa})}, \: \: \text{ where } \: \: \hat{k} = \sum_{p>q}^{M} \kappa_{pq} (\alpha_{p}^{\dagger} \alpha_{q} - \alpha_{q}^{\dagger} \alpha_{p})
    \end{aligned}
\end{equation}
is anti-Hermitian, $\hat{k}^\dag = -\hat{k}$. Note this places a restriction on the orbital rotation coefficient matrix, namely it is also anti-Hermitian, i.e. $\kappa_{pq} = -\kappa_{qp}^{*}$. Therefore, the rotation coefficients may be identified with a vector $\vec{\kappa}$ of length $\frac{M(M-1)}{2}$ (see Equation~\eqref{eq:TimSkew} in Appendix \ref{sec: BCH_fermionic}). By conjugating $H^{\text{ferm}}$ with the unitary $e^{-\hat{k}(\vec{\kappa})}$, the Hamiltonian is written in a new MO basis as:

\begin{equation}
    \begin{aligned} \label{eq:H_trans}
        H^{\text{ferm}} 
        \mapsto 
        &H^{\text{ferm}}(\vec{\kappa}) = e^{-\hat{k}(\vec{\kappa})} H^{\text{ferm}} e^{\hat{k}(\vec{\kappa})} \\
        = \sum_{mn}^{M}& \bar{h}_{mn}(\vec{\kappa})  \alpha_{m}^{\dagger} \alpha_{n} +  \sum_{mn\mu \nu}^{M} \frac{\bar{g}_{m\nu n \mu}(\vec{\kappa})}{2}  \alpha_{m}^{\dagger}  \alpha_{n}^{\dagger}  \alpha_{\mu} \alpha_{\nu}.
    \end{aligned}
\end{equation}

The bar above each coefficient is used as an identifier for each Hamiltonian coefficient that has undergone the orbital transformation. This change-of-basis is efficient to calculate on a conventional computer for any choice of $\vec{\kappa}$. To keep this text self-contained we detail this transformation in Appendix~\ref{sec: BCH_fermionic}. Note that some rotations coefficients (elements of $\vec{\kappa}$) can be unnecessary (or constrained) depending on the type of calculation being done. Post transformation, $H^{\text{ferm}}(\vec{\kappa})$ will be isospectral with $H^{\text{ferm}}$ as the map defined by Equation~\eqref{eq:U} is unitary. We highlight that if $\mathcal{\chi}$ is the set of fermionic operators appearing in $H$, then this set is closed under conjugation with the unitary transformation $U(\vec{\kappa})$. Operationally, this means only the coefficients of the Hamiltonian change under this transformation. 

In the Hartree-Fock procedure, the ground state is restricted to be a single Slater determinant, which by the variational principle will be an upper bound to the true ground state energy. The best such approximation is the ground state of the following  Hamiltonian:

\begin{equation}
    \begin{aligned} \label{eq:HF_H2}
        &H_{D}^{\text{ferm}}(\vec{\kappa}) 
        = \sum_{n}^{M} \bar{h}_{nn}(\vec{\kappa})  \alpha_{n}^{\dagger} \alpha_{n} + \\
        &\frac{1}{2} \sum_{\substack{m,n \\ m \neq n}}^{M}  \bigg(\bar{g}_{mmnn} (\vec{\kappa}) - \bar{g}_{mnnm}(\vec{\kappa}) \bigg) \alpha_{m}^{\dagger} \alpha_{m} \alpha_{n}^{\dagger} \alpha_{n}.
    \end{aligned}
\end{equation}
The subscript $D$ indicates the Hamiltonian is diagonal in the Fock basis defined by the creation and annihilation operators. Note the number of terms in Equation~\eqref{eq:HF_H2} scales quadratically as $\mathcal{O}(M^{2})$ where $M$ is the number of spin-orbitals/qubits. Appendix \ref{sec:ao2mo_HF} details how the two-electron integrals can be determined efficiently.

We re-iterate $H^{\text{ferm}}$ and $ H^{\text{ferm}}(\vec{\kappa})$ (Equation~\eqref{eq:H_trans}) share the same spectrum, but selecting only the diagonal terms (single Slater determinant restriction) results in the diagonal operators $H_{D}^{\text{ferm}}(\vec{\kappa})$ no longer being isopectral for different $\vec{\kappa}$. This is the crux of how Hartree-Fock optimizes the molecular orbital basis for a single Slater determinant.

Overall, the Hartree-Fock method is given by:
\begin{equation} \label{eq:HF_eq}
    \begin{aligned}
    E_{HF}^{\text{ferm}} &= \min_{\vec\kappa}  \big[ \bra{HF^{\text{ferm}}} e^{-\hat{k}(\vec{\kappa})} H^{\text{ferm}} e^{\hat{k}(\vec{\kappa})} \ket{HF^{\text{ferm}}} \big], \\
&= \min_{\vec\kappa}\big[ \bra{HF^{\text{ferm}}} H_{D}^{\text{ferm}}(\vec{\kappa}) \ket{HF^{\text{ferm}}} \big].
    \end{aligned}
\end{equation}
The corresponding ground state is the single slater determinant defined by:
\begin{equation} \label{eq:HF_state}
    \begin{aligned}
    \ket{HF^{\text{ferm}}} = \prod_{i\in \mathcal{O}_{\alpha} \cup \mathcal{O}_{\beta} } \alpha_{i}^{\dagger}\ket{vac}.
    \end{aligned}
\end{equation}
Where $\mathcal{O}_{\alpha}$ and $\mathcal{O}_{\beta}$ are the spin-orbital indices of occupied alpha (spin-up) and beta (spin-down) orbitals respectively. Usually $\mathcal{O}_{\alpha} \equiv \{2k ; k=1,2,
\dots, N_{\alpha}   \}$ and $\mathcal{O}_{\beta} \equiv \{2k+1; k=1,2,
\dots, N_{\beta}   \}$, where $N_{\alpha}$ and $N_{\beta}$ are the number of spin-up and -down electrons respectively. The energy is minimized with respect to $\vec{\kappa}$, typically by gradient approaches that use the electronic orbital gradient \cite[eq 10.8.18]{helgaker2013molecular} and electronic Hessian \cite[eq 10.8.53]{helgaker2013molecular}. The resulting MOs from Equation~\eqref{eq:HF_state} may not be in canonical form\footnote{In mathematics a canonical matrix is a member of an equivalence class of matrices that have a specific form. The canonical Fock matrix is obtained from a particular choice of molecular orbitals (the basis) that results in the Fock matrix being diagonal \cite{glaesemann2010ordering}.}, but this can be remedied with a single diagonalization (of the Fock matrix described in atomic orbital basis) at the conclusion of the optimization \cite{wong1995approaches}. 

To summarize, the Hartree-Fock algorithm optimizes the molecular orbital basis of the problem with respect to a single Slater determinant. It achieves this via a specific change of basis (Equation~\eqref{eq:U}) and minimizes the energy with respect to this basis transformation. Another way of interpreting Equation~\eqref{eq:HF_eq} is by the Thouless theorem \cite{thouless1960stability}, which says the transformation of a single
Slater determinant by $e^{\hat{k}(\vec{\kappa})}$ will also be a (non-orthogonal) single Slater determinant. 

To map the HF Hamiltonian to a spin problem we re-write Equation~\eqref{eq:HF_H2} under the Jordan-Wigner transform as:
\begin{equation}\label{eq:HF_H3}
    \begin{aligned}
        H_{D}(R) ={} & \sum_{n}^{M} \frac{\bar{h}_{nn}(R)}{2} \bigg(I - Z_{n}\bigg) +\\
        \sum_{\substack{m,n \\ m \neq n}}^{M} &\frac{\bar{g}_{mmnn}(R)- \bar{g}_{mnnm}(R)}{8} \bigg(I - Z_{n} - Z_{m} + Z_{m}Z_{n} \bigg) \\
        ={} & \sum_{j=1}^{1+\frac{M(M+1)}{2}} c_{j}(R) P_{j}.
    \end{aligned}
\end{equation}

Here, $R$ is a placeholder for $\vec{\kappa}$ and the molecular orbital coefficient matrix $C$. Note the tensor product symbol between Pauli operators is implied and the trivial identity factors acting on the other qubits are omitted for brevity. Each $c_{j}(R) \in \mathbb{R}$ is a real coefficient. In this encoding, the diagonal Pauli Hamiltonian will only be composed of $n$-fold tensor products $P_j$ consisting of $I,Z$ that act non-trivially on at most two qubits. This should be clear from the fact the Fermionic Hamiltonian in Equation~\eqref{eq:HF_H2} is composed of at most a product of two number operators. The Pauli Hamiltonian in Equation~\eqref{eq:HF_H3} therefore describes a Quadratic Unconstrained Spin Optimization (QUSO) problem by definition.  Note the Hartree-Fock state of Equation~\eqref{eq:HF_state} under the Jordan-Wigner transformation is still a single computational basis state. 

As with all optimization problems, convergence onto local minima or saddle points is a failure mode for the algorithm. In SCF electronic structure calculations performing internal stability calculations avoids convergence onto saddle points and excited states within the specified search space. External stability calculations relax different constraints on the search space. Examples include allowing imaginary molecular orbitals (complex RHF) or allowing alpha and beta spin-orbitals to occupy different spatial orbitals (unrestricted Hartree-Fock). Such relaxations can be used to determine different instabilities, for example singlet-triplet instabilities. We do not consider external instabilities in this work and instead focus on internal stabilities. 

At a high level, internal stabilities are determined via a characteristic set of matrices. For a calculation to be internally stable, all eigenvalues of particular linear combinations of the  characteristic matrices must have positive semidefinite character (all eigenvalues are real). A negative eigenvalue of the characteristic problem indicates an internal instability. In other words, another single Slater determinant must exist with a lower  expectation value. See \cite{vcivzek1967stability, paldus1970stability, seeger1977self, bauernschmitt1996stability, dehareng2000hartree, sharada2015wavefunction} for in-depth discussions on stability analysis. 

In all single-Slater-determinant SCF methods, the essential goal is to find a molecular orbital basis in which that single determinant has the lowest energy. In the context of this paper, each method will output a converged $\vec{\kappa}$ along with a single Fock (or computational basis) state. Practically, most chemistry codes will not provide this vector $\vec{\kappa}$. Instead, they usually only output the optimized molecular orbitals (as the coefficient matrix $C$) and the single Fock state. Mathematically, these are equivalent, as they each provide the information required to calculate the one- and two-electron integrals ($h_{pq}$ and $g_{pqrs}$) and subsequently construct the full electronic molecular Hamiltonian $H$ in Equation~\eqref{eq:spin_free_H}. We remark that all orbital localization schemes, for example natural orbitals, again just define a particular molecular orbital basis and thus are captured by the analysis here. 

This is also true for Kohn-Sham density functional theory (KS-DFT) \cite{kohn1965self}, which uses the electron density to determine the ground state energy via the use of functionals. The Kohn-Sham method introduces a fictitious system of non-interacting particles that generate the same density as the system of interacting particles. Kohn-Sham DFT is formally exact; however, the explicit form of the exchange-correlation functional is unknown and thus is approximated. The energy obtained is still variational, but with respect to a model Hamiltonian and not the true system Hamiltonian.  Fortunately, the converged KS-DFT molecular orbitals are still a valid MO basis and thus the true molecular Hamiltonian can be written in this basis. This means a variational QUBO problem can be defined with respect to the true Hamiltonian, just as for the other SCF methods.

In this Section we demonstrated that Hartree-Fock consists of both continuous and discrete optimization. The optimization of a discrete function naturally introduces the concept of pseudo-Boolean functions, as discussed in the following Section \ref{sec:QUBO}.

\subsection{Quadratic Optimization}\label{sec:QUBO}

A pseudo-Boolean function is a real-valued function $f(\vec{x}) = f(x_{1}, x_{2}, \hdots, x_{n})$ of $n$ binary variables that is a mapping from $\vec{x} \in \{0,1\}^{n} \mapsto \mathbb{R}$ \cite{boros2020compact}. Pseudo-Boolean optimization (PBO) problems seek to minimize such nonlinear functions:

\begin{equation}
    \text{min} \big\{ f(\vec{x}): \vec{x} \in \{0,1\}^{n} \big\}.
\end{equation}

Quadratic unconstrained binary optimization (QUBO) problems are an optimization over pseudo-Boolean functions that are limited to contain at most quadratic terms. These are commonly written as:


\begin{equation} \label{eq:QUB)}
    \begin{aligned}
    g(\vec{x}) &= \vec{x}^{T} Q \vec{x} + \vec{b}{\:}^{T}\vec{x} + c, \\
    &= \sum_{ij} Q_{ij} x_{i}  x_{j} + \sum_{i} b_{i} x_{i} + c.
    \end{aligned}
\end{equation}
Here the quadratic operator is represented by the matrix $Q \in \mathbb{R}^{n \times n}$, the vector $\vec{b} \in \mathbb{R}^{n}$ gives the linear terms and a constant given by $c \in \mathbb{R}$. Here $g(\vec{x})$ is the pseudo-Boolean function of $n$ binary variables. Note for binary variables $x_{i}x_{i} = x_{i}$ and thus sometimes all terms (excluding the constant) are included  in $Q$.

A binary optimization can be mapped to a spin optimization (and \textit{vice versa}) as follows:
\begin{equation} \label{eq:mapping_vars}
    x_{i}  \longleftrightarrow \frac{1 - z_{i}}{2},
\end{equation}
where an optimization over binary variables $x_{i} \in \{0, 1 \}$ has been replaced by an optimization over spin variables $z_{i} \in \{+1, -1 \}$. In general such a spin problem is known as a Polynomial Unconstrained Spin Optimization (PUSO) problem. 

The MaxCut problem is one of the most well-known optimization problems and is a specific instance of a QUSO/QUBO problem. It is a combinatorial optimization problem in graph theory where one needs to partition the vertices of a weighted graph into two distinct sets, such that the sum of the weights of the edges connecting those two sets is maximized. Essentially, the problem is to find the "cut" (partition) through the graph that has the highest total edge weight between the two sets. The approximation ratio for the MaxCut is the factor by which an heuristic solution differs from the optimal solution. See Appendix \ref{sec:MaxCut} for further information.

Applying the semidefinite programming algorithm of Goemans-Williamson (GW) \cite{goemans1994879} to MaxCut guarantees an approximation ratio of at least 0.878 compared to the optimal cut. In other words, it guarantees a solution that is at least $87.8$ \% as good as the optimal solution. However, this requires the graph to have non-negative edge weights \cite{goemans1994879}. Finding an algorithm that improves this ratio has major implications for the Millennium Prize Problem of whether $\mathrm{P} = \mathrm{NP}$. This relates to the unique games conjecture \cite{khot2002power}.  Khot, Kindler, Mossel and O'Donnel \cite[Theorem 1]{khot2007optimal} show that if the unique games conjecture is true, then the Goemans-Williamson approximation algorithm for MaxCut is optimal unless $\mathrm{P} = \mathrm{NP}$.

\section{SCF model description} \label{sec:SCF_model}

In this section, we derive the SCF models proposed in this work, highlighting and discussing key parts of each routine.  For brevity, Appendix  \ref{sec:MaxCutChem} outlines the mapping of the problem first to a QUSO, followed by to QUBO (Appendix \ref{sec:QUSO2QUBO}), and then shows how the QUSO is mapped to a MaxCut problem (Appendix \ref{sec:QUSO2MAXCUT}). First, in Section \ref{sec:SCF_opt_new}, we introduce the core of our SCF algorithm (QUSO-SCF). 
In Section \ref{sec:HF-maxcut} we discuss how the optimization problem can be solved by solving a series of graph problems.  Section \ref{sec:QUBO-HF} then discusses how our new model can be used to improve conventional SCF calculations and discuss performance guarantees. 

\begin{figure*}[ht]
    \centering
    \begin{subfigure}[b]{0.45\textwidth}
\begin{tikzpicture}[node distance=2cm]
\node (start) [startstop] {Start};
\node (in1) [io, below of=start, xshift=0.0cm] {Input: $\vec{\kappa}_{0}$, ${h}_{pp}$, $g_{ppqq}$, $g_{pqqp}$ };
\node (pro1) [process, below of=in1, xshift=0cm] {Calculate $\bar{h}_{nn}(\vec{\kappa})$, $\bar{g}_{mmnn}(\vec{\kappa})$ and $\bar{g}_{mnnm}(\vec{\kappa})$};
\node (pro2) [process, below of=pro1, xshift=-2.5cm] {$H_{D}^{\text{ferm}}(\vec{\kappa})$};
\node (pro3) [process, right of=pro2, xshift=2.5cm] {Build $H_{D}(\vec{\kappa})$};
\node (pro35) [process, below of=pro3, yshift=0.35cm] {$H_{\text{QUSO}}(\vec{\kappa}) $ / $ H_{\text{QUBO}}(\vec{\kappa})$ / $ H_{\text{MaxCut}}(\vec{\kappa})$};
\node (pro4) [process, left of=pro35, , xshift=-2.0cm, yshift=-0.15cm] {Solve: $\ket{b_{\min}}$ (Ising machine / classical heuristic / quantum computer)};

\node (dec1) [decision, below of=pro4, xshift=2cm] {$E(\ket{b_{\min}}; \vec{\kappa})$ converged?};
\node (stop) [startstop, below of=dec1, xshift=2cm] {Finished};
\node (pro5) [process, below of=dec1, xshift=-1.8cm] {Update $\vec{\kappa}$};

\draw [arrow] (start) -- (in1);
\draw [arrow] (in1) -- (pro1);
\draw [arrow, dashed] (pro1) -- (pro2);
\draw [arrow] (pro1) -- (pro3);
\draw [arrow, dashed] (pro2) -- (pro3);
\draw [arrow] (pro3) -- (pro35);
\draw [arrow] (pro35) -- (pro4);
\draw [arrow] (pro4) -- (dec1);

\draw [arrow] (dec1.south west) -- node[anchor=east] {no} (pro5);
\draw [arrow] (dec1.south east) -- node[anchor=west] {{ }  yes} (stop.north);
\draw[arrow] (pro5) -- +(-2.5,0) |- (pro1);
\end{tikzpicture}
\caption{Algorithm \ref{alg:cap1}} \label{fig:alg_overview}
\end{subfigure}
    \hfill
    \hspace{0.05\textwidth}
    \begin{subfigure}[b]{0.45\textwidth}
    \begin{tikzpicture}[node distance=2cm]
    \node (start) [startstop] {Start};
    \node (in1) [io, below of=start, xshift=0.0cm] {Input: $\gamma_{i=0}$};
    \node (pro1) [process, below of=in1, xshift=0cm] {Build Fock matrix: $F(\gamma_{i})$};
    \node (pro2) [process, below of=pro1, xshift=-2.5cm] {Solve: $F(\gamma_{i}) C_{i} = SC_{i}e$ defining MOs: $C_{i}^{opt}$};
    \node (pro3) [process, right of=pro2, xshift=2.5cm] {Build $H_{D}^{\text{ferm}} (C_{i}^{opt})$};
    \node (pro35) [process, below of=pro3, yshift=0.3cm] {$H_{\text{QUSO}} $ / $ H_{\text{QUBO}}$ / $ H_{\text{MaxCut}}$};
    \node (pro4) [process, left of=pro35, , xshift=-2.0cm, yshift=-0.2cm] {Solve: $\ket{b_{\min}}$ (Ising machine / classical heuristic / quantum computer)};

    \node (dec1) [decision, below of=pro4, xshift=2cm] {$E(\ket{b_{\min}})$ converged?};
    \node (stop) [startstop, below of=dec1, xshift=2cm] {Finished};
    \node (pro5) [process, below of=dec1, xshift=-1.8cm] {$\gamma_{i+1} = C_{i}^{opt}T\big(\ket{b_{\min}} \big) C_{i}^{opt})^{\dagger}$};
    
    \draw [arrow] (start) -- (in1);
    \draw [arrow] (in1) -- (pro1);
    \draw [arrow] (pro1) -- (pro2);
    \draw [arrow, dashed] (pro2) -- (pro3);
    \draw [arrow, dashed] (pro3) -- (pro35);
    \draw [arrow] (pro2) -- (pro35);
    \draw [arrow] (pro35) -- (pro4);
    \draw [arrow] (pro4) -- (dec1);
    
    \draw [arrow] (dec1.south west) -- node[anchor=east] {no} (pro5);
    \draw [arrow] (dec1.south east) -- node[anchor=west] {{ }  yes} (stop.north);
    \draw[arrow] (pro5) -- +(-2.5,0) |- (pro1);
    \draw [draw, xshift=-6.5cm, yshift=-11.cm, xshift=2cm] -- node {$i=i+1$} (pro4);
    \end{tikzpicture}
    \caption{Algorithm \ref{alg:cap2}} \label{fig:alg_overview2}
    \end{subfigure}
    \caption{Outline of the new SCF algorithms derived in this work. (a) Algorithm \ref{alg:cap1} is based off the optimization defined by equation  \ref{eq:HF_qUso}. Here $\vec{\kappa}$ can be updated with the orbital gradient and Hessian, note the Fock matrix is not diagonalized in this routine. (b) Algorithm \ref{alg:cap2} is based off the optimization defined by Equation \eqref{eq:traditional}. Here $\gamma_{i}$ is the standard one-particle density matrix in the AO representation and $T(\ket{b_{\min}})$ is the spin-free one-particle density matrix in the spatial MO basis. To build and solve (diagonalize) the Fock matrix see standard textbooks \cite{szabo2012modern, helgaker2013molecular}. See Appendix \ref{sec:pseudo_algs} for full pseudo code for each approach. Dashed lines indicate the Fermionic Hamiltonian construction followed by Jordan-Wigner mapping that can be circumvented for the most efficient implementations.}
    \label{fig:algorithms}
\end{figure*}
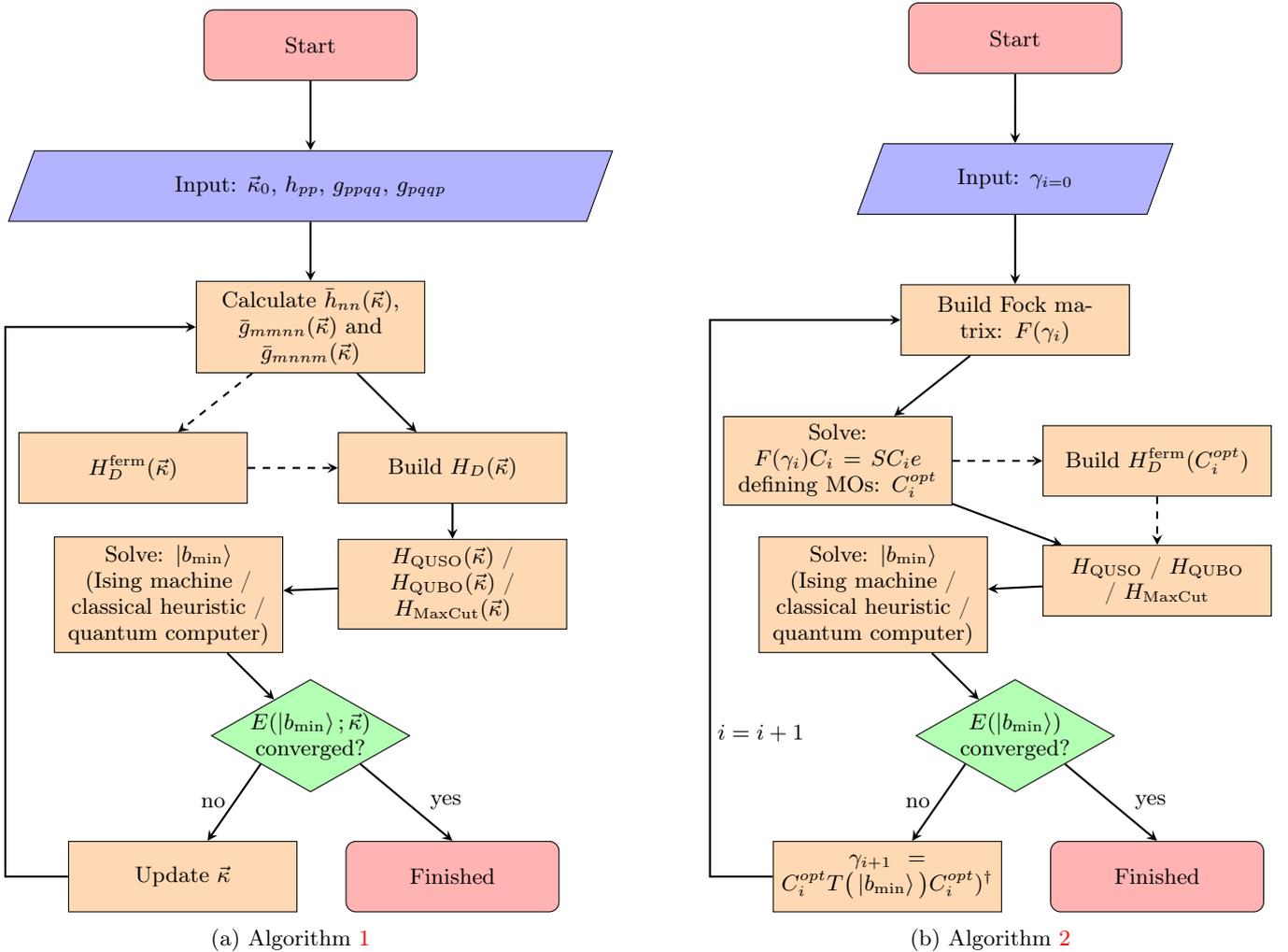

\subsection{SCF via quadratic optimization} \label{sec:SCF_opt_new}

We now define two new SCF procedures that differ slightly from conventional Hartree-Fock.  Each is introduced as a QUSO problem. The QUBO and MaxCut versions of these algorithms follow in exactly the same way, except the Hamiltonian is modified. We provide these mappings in Appendix \ref{sec:MaxCutChem}. The complete pseudo-algorithms for both approaches can be found in the Appendix \ref{sec:pseudo_algs}. The two algorithms differ in how they specify changes in the molecular orbital basis. The basis of molecular orbitals may be described in terms of either the molecular orbital coefficient matrix $C$, or the vector of parameters $\vec\kappa$, introduced in Equation \eqref{eq:U}. The molecular orbital coefficient matrix $C$ is a matrix of coefficients that relates a set of molecular orbitals to a fixed basis of atomic orbitals. The vector of parameters $\vec\kappa$ defines a unitary that maps the molecular orbitals, defined by a fixed molecular orbital coefficient matrix $C$, to a new set of orbitals. These two approaches correspond to different ways of exploring the space of orbital bases for a given molecular problem. For any $\vec\kappa$ there will be an equivalent molecular orbital coefficient matrix $C$ realizing the same orbitals. However, the structure of the optimization problem will differ between Algorithm $1$ and Algorithm $2$. The comparative advantage of different optimization problems depends on problem structure~\cite{NoFreeLunch}, and hence it is useful to define Algorithm $1$ and $2$ separately.

\subsubsection{Algorithm $1$}

The first optimization problem is defined as:
\begin{equation} \label{eq:HF_qUso}
    \begin{aligned}
    E_{\text{Alg-}1}^{\text{min-eig}} &= \min_{\vec\kappa} \biggl[ \min_{\ket{b} \in \mathcal{S} \subset \mathcal{D}} \big[ \bra{b} H_{D}(\vec{\kappa}) \ket{b} \big] \bigg]. 
    \end{aligned}
\end{equation}
Here, optimization of the molecular orbital basis is performed in terms of the parameter vector $\vec\kappa$; what is being optimized in the continuous part of the problem is a unitary matrix defining a map to new molecular orbitals.

In the above, $\mathcal{S}$ is the discrete set of all single computational basis states that respect the symmetry of the underlying electronic structure problem (Slater determinants), for example each element in the set has the correct number of alpha/beta electrons and desired spin multiplicity. This is therefore a Quadratic Constrained Spin Optimization (QCSO) problem. $\mathcal{S}$ is a subset of $\mathcal{D}$, which is the set of all computational basis states (all possible Fock states) and thus has size $2^{M}$. In implementations of this algorithm one may not be able to explicitly restrict optimization to $\mathcal{S}$, for example in Quantum Annealing. In such cases, we later address in Section~\ref{sec:benefits_and_challenges} and Appendix~\ref{sec:mol_sym} how to bias optimization towards elements of $\mathcal{S}$ from the full set of Fock states $\mathcal{D}$. Note an optimization over $\mathcal{D}$ defines a quadratic unconstrained spin optimization (QUSO) problem.

In Equation \eqref{eq:HF_qUso}, the inner optimization is trying to find the best computational basis state (Fock state) that gives the lowest energy in the current molecular orbital basis defined by $\vec{\kappa}$. The outer loop optimizes the molecular orbital basis, i.e. $\vec{\kappa}$, with respect to the energy. The optimization is concluded when the energy reaches a minimum with respect to $\vec{\kappa}$. Figure \ref{fig:alg_overview} illustrates each step of this approach which we denote Algorithm \ref{alg:cap1}.

Note this construction differs from standard Hartree-Fock, as the HF reference state is not fixed during optimization over $\vec{\kappa}$ and instead an inner optimization over different computational basis states is performed. For each MO basis a QUBO optimization is done to find the optimal single Slater determinant described in that basis, which is then used to optimize $\vec{\kappa}$. 

\subsubsection{Algorithm $2$}

The second algorithm is formulated as:
\begin{equation} \label{eq:traditional}
    \begin{aligned}
    E_{\text{Alg-}2}^{\text{min-eig}} &=  \min_C \bigg[  \min_{\ket{b} \in \mathcal{S} \subset \mathcal{D}} \big[ \bra{b} H_{D}(C) \ket{b} \big] \bigg],
    \end{aligned}
\end{equation}
where
\begin{equation}
    \begin{aligned} \label{eq:HF_HC}
        &H_{D}^{\text{ferm}}(C) 
        = \sum_{n}^{M} \bar{h}_{nn}(C)  \alpha_{n}^{\dagger} \alpha_{n} + \\
        &\frac{1}{2} \sum_{\substack{m,n \\ m \neq n}}^{M}  \bigg(\bar{g}_{mmnn} (C) - \bar{g}_{mnnm}(C) \bigg) \alpha_{m}^{\dagger} \alpha_{m} \alpha_{n}^{\dagger} \alpha_{n},
    \end{aligned}
\end{equation}
is the diagonal (with respect to a Fock basis) part of Equation \eqref{eq:spin_free_H}. Here, the optimization is performed over the molecular orbital coefficient matrix $C$, so what is being optimized is the map from the fixed basis of atomic orbitals to the molecular orbitals.

In Algorithm \ref{alg:cap2}, for a particular $C$, the Fock matrix is diagonalized to obtain a new $C$, which defines a new Fock matrix. This repeated diagonalization process of the Fock matrix is performed until the Fock matrix (and therefore $C$) converges. Figure \ref{fig:alg_overview2} depicts each step of this method. This approach differs from conventional Hartree-Fock implementations in that the Fock matrix has dependence on $C$ and the one-particle reduced density matrix ($1$-RDM). The $1$-RDM is fixed by a specific Slater determinant (Equation \eqref{eq:HF_state}) in the MO basis, rather than being optimized at each step. 

In Equation \eqref{eq:traditional}, for an initial $C$ the Hamiltonian is defined as in Equation \eqref{eq:HF_HC}. A combinatorial optimization over all Slater determinants in this basis is then performed to find the best possible reference state in the current molecular orbital basis. This single Slater determinant will define a particular $1$-RDM. Along with the current $C$ matrix, this can be used to construct a new Fock matrix, which is subsequently diagonalized to obtain a new $C$. A new Hamiltonian (Equation \eqref{eq:HF_HC}) is then written and this process is repeated until the energy converges. 

The difference between the Algorithm $2$ and traditional HF lies in the fact that the one-particle reduced density matrix in Equation \eqref{eq:traditional} is not fixed by a particular Slater determinant and instead optimized over. The flexibility to choose a lower energy single Slater determinant, as the molecular orbital basis is being optimized, is exactly why this approach should converge faster and be more resistant to local minima.

\subsubsection{Benefits and Challenges of Algorithms $1$ and $2$}\label{sec:benefits_and_challenges}

When applying the Hartree-Fock procedure, it is well-known that convergence often leads to local minima \cite{van2006starting, vaucher2017steering, lehtola2019assessment}. An advantage of these newly proposed SCF approaches is they should be more resistant to local minima as the single reference Fock state is allowed to change rather than being fixed.

Changing the Slater determinant during the optimization also introduces new challenges.
While the optimal $\vec{\kappa} / C$ and $ \ket{b}$ will give the true minimum eigenvalue of the QCSO problem (and $H_{D}^{\text{ferm}}(\vec{\kappa})$), solving the QUSO problem (over $\mathcal{D}$ rather than $\mathcal{S})$ can provide solutions with incorrect symmetries, whereas a fixed HF reference state can be preselected to be in the correct symmetry sector. For example, solutions with the wrong number of electrons can be obtained via this approach (i.e. a solution in the wrong Fock sector). Fortunately, it is easy to check these conditions as $ \ket{b}$ is a single computational basis state (representing a single Slater determinant).  In Appendix~\ref{sec:mol_sym}, we show how the unconstrained spin problem can be modified such that problem symmetries are accounted for. In summary, the problem is reformulated as:
\begin{equation} \label{eq:HF_vp_QUSO}
    \begin{aligned}
    E_{QUSO}^{V_{P}} &= \underset{R}{\mathrm{min}}  \bigg[ \underset{\ket{b} \in \mathcal{D}}{\mathrm{min}} \big[ \bra{b} H_{D}(R) + \lambda V_{P} \ket{b} \big] \bigg].
    \end{aligned}
\end{equation}
Here $V_{P}$ penalizes any Slater determinant not in the correct symmetry sector by $\lambda$, which takes a large positive value.  This provides a way to favor certain molecular symmetries during the combinatorial optimization. 

In the following subsections we describe various approaches, both quantum and classical, to the inner discrete optimization loop in Algorithms \ref{alg:cap1} and \ref{alg:cap2}. In Sections~\ref{sec:HF-maxcut} and~\ref{sec:QUBO-HF} we describe a mapping to the problem MaxCut and give performance guarantees arising from semidefinite programming approaches to MAXCUT. In Section~\ref{sec:Qalgs} we describe four quantum approaches to this optimization, arising from Grover Search, QAOA, Quantum Annealing and Decoded Quantum Interferometry (DQI).  

\subsection{MaxCut SCF} \label{sec:HF-maxcut}

A QUSO problem can be transformed into an equivalent MaxCut problem using a single auxiliary bit. In the interest of keeping this paper self-contained we provide the procedure in Appendix \ref{sec:MaxCutChem}; in particular, see Equation \eqref{eq:maxcut_full}. The resulting graph problem in general will contain negative edge weights. The problem of solving MaxCut with negative weights is discussed in \cite{nesterov1998semidefinite, charikar2004maximizing, goemans1995improved}. The problem can be written via a graph Laplacian, which is a symmetric real matrix and a $0.56$ approximation ratio algorithm is always possible \cite{alon2004approximating, charikar2004maximizing}. Furthermore, if this Laplacian matrix is positive semidefinite, then a $\frac{2}{\pi}\approx 0.6$ approximation algorithm is possible \cite[Theorem 3.3]{nesterov1998semidefinite} \cite{charikar2004maximizing} Alternatively, the Goemans-Williamson result can be generalized to negative weight graphs \cite[Section 3.2]{goemans1995improved}. This could provide a tighter bound on the approximation ratio than $0.56$ or $\frac{2}{\pi}$. However, this bound is problem-dependent as the negative edge weights are defined by the particular problem (see Appendix \ref{sec:MaxCut}, specifically Equation \eqref{eq:neg_weights2}).

We remark here that if the orbitals are real, then the two-body integrals ($\bar{g}_{mmnn}$, $\bar{g}_{mnnm}$) will be positive definite \cite[Section 9.12.4]{helgaker2013molecular} \cite{roothaan1951new}. Furthermore, the diagonal elements are positive \cite[eq (9.12.24)]{helgaker2013molecular}. Thus, when defining a MaxCut problem the number of negative edge weights will at worst scale linearly. The positive edge weights will scale quadratically due to the $\big(\bar{g}_{mmnn} (\vec{\kappa}) - \bar{g}_{mnnm}(\vec{\kappa}) \big) \geq 0 \; \forall m,n$ terms being all positive. Appendix \ref{sec:MaxCutChem} provides a richer analysis of this. Thus, when writing down the MaxCut optimization problem the number of positive edge weights will be much greater than the number of negative edge weights. Importantly, if penalty terms are utilized then further negative weights can occur, but this will again be problem-dependent. Note such modifications can also remove negative edges from the problem.

The MaxCut-SCF approach to single reference SCF methods has the same complexity as Hartree-Fock and Kohn-Sham Density Function Theory, which are both NP-complete problems in general \cite{whitfield2014np, whitfield2014computational}. However, applying semi-definite programming optimization methods to the MaxCut-SCF problem ensures a solution that is at same fraction of the optimal solution at each step. In particular, at each step of both algorithms, while the MO basis is being optimized the $1$-RDM considered remains within a certain fraction of the optimal one, determined by the specific approximation algorithm used. This guarantee is not present in conventional Hartree-Fock approaches. MaxCut-SCF can therefore be used to either update the MO basis at each step of an SCF routine or check if a better solution is possible in a specified MO basis (i.e. it can find a lower energy Fock state than the SCF method determines).

We remark here that this analysis does not hold for multi-reference SCF methods, such as Complete Active Space self-consistent field (CASSCF) \cite{roos1980complete}, as the ground state is no longer approximated by a single Fock state.

\subsection{QUBO/MaxCut Boosted SCF} \label{sec:QUBO-HF}

To improve any SCF result using the QUBO/MaxCut formulation, first the diagonal terms of the full molecular Hamiltonian are written in the specified basis, i.e. $H_{D}^{\text{ferm}}(\vec{\kappa})$  in Equation \eqref{eq:HF_H2}. This is then mapped to a QUBO or MaxCut problem. This defines an optimization problem with the goal of finding the lowest energy Slater determinant in the MO basis, defined by the conventional SCF technique.

There are two major benefits in finding a better single Slater determinant: (1) The solution provides a better upper bound to the true ground state energy by the variational principle and (2) It has major implications for single reference post Hartree-Fock methods, where essentially these can all be started from a better reference point.

\subsection{Hybrid Quantum-Classical SCF Algorithms} \label{sec:Qalgs}

In this subsection, we outline four hybrid quantum-classical SCF algorithms based on Grover Adaptive Search, QAOA, Quantum Annealing and DQI. We note all these algorithms differ from the Hartree-Fock Variational Quantum eigensolver (VQE) technique performed by Google AI Quantum and collaborators \cite{google2020hartree} on their quantum computer. In their approach, the input state is fixed and orbital rotations  of the form in Equation \eqref{eq:U} were performed on the quantum device. A critical perspective on this work is that the $1$- and $2$-RDMs are entirely determined by their initial state prior to orbital rotations. Since the initial state was fixed the orbital updates could have been obtained classically, for instance, using the orbital gradient and Hessian \cite{helgaker2013molecular}.

In the following subsections, for each new quantum-SCF algorithm, the ``input" Slater determinant is changing and thus while it is efficient to determine the $1$- and $2$-RDMs from the chosen state, selecting the optimal Slater determinant is the hard part of the problem. Each of the quantum SCF methods can therefore be thought of as using a different quantum algorithm to choose this state (single Slater determinant). The orbital updates are then determined classically, as this step is efficient to do on conventional computers. To capture this, we annotate Equation \eqref{eq:HF_vp_QUSO}: 
\begin{equation} \label{eq:HF_vp_QUSO_quantum}
    \begin{aligned}
    E_{QUSO}^{V_{P}} &= \underbrace{\min_R}_{\substack{\text{classical}\\\text{algorithm}}}  \bigg[ \underbrace{\min_{\ket{b} \in \mathcal{D}} \big[ \bra{b} H_{D}(R) + \lambda V_{P} \ket{b}}_{\substack{\text{quantum}\\ \text{algorithm}}} \big] \bigg],
    \end{aligned}
\end{equation}
which encapsulates the core concept of all the new quantum SCF techniques introduced next. In contrast to the VQE approach~\cite{google2020hartree}, the inner optimization was not performed (i.e. $\ket{b}$ was fixed in their method), and only the outer optimization (over $R$) was performed, which they did on their quantum device.

\subsubsection{GAS-SCF} 
Gilliam, Woerner, and Gonciulea \cite{gilliam2021grover} recently demonstrated that quadratic unconstrained binary optimization (QUBO) problems can be accelerated using Grover Adaptive Search (GAS) \cite{bulger2003implementing, baritompa2005grover}.  The main idea of this approach is to iteratively apply the Grover search algorithm to find the optimum value of an objective function. This is achieved by flagging all values smaller than a threshold in order to find a better solution. This provides a quadratic speed-up for QUBO problems compared to brute force search. 

Since the inner optimization in Equation \eqref{eq:HF_vp_QUSO_quantum} is equivalent to a QUBO problem (see Appendix \ref{sec:QUSO2QUBO}), the GAS approach can be leveraged in our setting and thus providing a quadratic speed-up. We denote this quantum approach as GAS-SCF. It is worth noting that Grover-like speed-ups are typically realized only in the asymptotic regime. For practical problem sizes, constant overheads in an algorithm’s implementation may negate any theoretical advantage. A detailed investigation of GAS-SCF is left for future work, as it lies beyond the scope of this paper.

\subsubsection{QAOA-SCF} 
QAOA, or the Quantum Approximate Optimization Algorithm \cite{farhi2014quantum}, has emerged as a prominent strategy for solving combinatorial optimization problems on near-term quantum hardware. In the context of this work, a QAOA-SCF algorithm can be written as:

\begin{equation} \label{eq:QAOA}
    \begin{aligned}
    E_{QAOA\text{-}SCF} &= \min_R \bigg[ \min_{\vec{g}, \vec{b}} \big[ \bra{\vec{g}, \vec{b}} H_{c}(R) \ket{\vec{g}, \vec{b}} \big] \bigg]
    \end{aligned}
\end{equation}
where the QAOA anstaz is defined as:
\begin{equation} \label{eq:QAOA_ansatz}
    \begin{aligned}
    \ket{\vec{g}, \vec{b}} &= \prod_{k=l}^{1} e^{-ib_{k}H_{m}} e^{-ig_{k}H_{c}(R)} \ket{s} \\
    &= e^{-ib_{l}H_{m}} e^{-ig_{l} H_{c}(R)} \hdots  e^{-ib_{1} H_{m}} e^{-ig_{1} H_{c}(R)} \ket{s}.
    \end{aligned}
\end{equation}
$H_{c}(R) = H_{D}(R) + \lambda V_{P}$ is the "cost" Hamiltonian (equation  \ref{eq:HF_vp_QUSO_quantum}) and $H_{m}$ is the "mixer" Hamiltonian, while $l$ denotes the number of layers. The initial state $\ket{s}$ is usually the uniform superposition over computational basis states, which can be achieved by applying a Hadamard gate on each qubit: $\ket{s} = H^{\otimes n}\ket{0} = \frac{1}{\sqrt{2^{n}}}\sum_{k=0}^{2^{n}-1} \ket{k}$.

Note that the implementation of time evolution under the “cost” Hamiltonian incurs no Trotter error, as all terms in the Hamiltonian commute. For details on the implementation and variations of the QAOA algorithm, refer to the recent review \cite{blekos2024review} and references therein. Detailed investigation of QAOA-SCF is left for future work.

Finally, we note that if a more general anstaz were utilized then the inner optimization will be solved by the Variational Quantum Eigensolver (VQE) algorithm \cite{peruzzo2014variational}, hence VQE-SCF is obtained by this approach.

\subsubsection{QA-SCF} 
Quantum Annealing (QA) is an innovative form of analog computation designed to utilize quantum mechanical effects to search for the optimal solutions of Ising problems.  The inner optimization of Equation \eqref{eq:HF_vp_QUSO_quantum} defines an Ising problem. Therefore,  the QA-SCF algorithm simply solves the inner optimization via Quantum Annealing, followed by updating the molecular orbitals (outer optimization) through classical optimization.

\subsubsection{DQI-SCF} 
Decoded Quantum Interferometry (DQI) is a quantum algorithm designed to solve combinatorial optimization problems by finding approximate solutions \cite{jordan2024optimization}. It exploits the sparse Fourier spectrum of specific objective functions, reducing the optimization task to a decoding problem. 

The MaxCut SCF problem of Equation \eqref{eq:maxcut_full} can be converted to a weighted MAX-2-XORSAT instance as:

\begin{equation} \label{eq:QDI_alg}
    \underset{\vec{z}}{\mathrm{max}}  \bigg[ \sum_{ij}A_{ij} \frac{I-z_{i}z_{j}}{2} \bigg] \mapsto \underset{\vec{x}}{\mathrm{max}}  \bigg[ \sum_{ij}A_{ij} \big(x_{i}\oplus x_{j} \big) \bigg],
\end{equation}
where each clause is simply:
\begin{equation}  
    x_{i}\oplus x_{j}=\begin{cases}
1, & \text{ if } A_{ij}>0,\\
0, & \text{ if } A_{ij}<0
\end{cases}.
\end{equation}
The goal is to maximize the total satisfied XOR clauses, which will maximize the cost function. This cost function can be approximately solved using the DQI algorithm. 

We note that the right-hand side of Equation \eqref{eq:QDI_alg} may be able to be modified to include more constraints. For example, imposing other molecular symmetries, such as alpha/beta spin parity. Another example is enforcing $\langle S^{2} \rangle=1$, which requires quartic terms. Conventional SCF approaches also apply such constraints to reduce the complexity of the problem. In general, these modifications will give MAX-$k$-XORSAT problems. This could improve the results obtained from DQI-SCF. We leave an investigation of DQI-SCF to future work including whether a QUSO/QUBO mapping is possible (removing the single ancilla qubit).
\section{Numerical Study\label{sec:discussion}} 
The \texttt{\texttt{PySCF}} package \cite{sun2020recent} was used to obtain the molecular integrals in order to construct each molecular Hamiltonian. For the  \ce{OH-} study we used the 6-31G basis set \cite{hehre1972self},  whereas the \ce{N2} study utilized the Dunning correlation-consistent polarized valence double/triple/quadruple zeta (cc-pVDZ, cc-pVTZ, cc-pVQZ) basis sets \cite{dunning1989gaussian}. Just as in restricted Hartree-Fock (RHF), the alpha and beta spatial orbitals were restricted to be identical for all calculations in this work.  The Jordan-Wigner transformation \cite{wigner1928paulische} was used to convert the fermionic Hamiltonian of Equation \eqref{eq:HF_H2} into a qubit Hamiltonian as per Equation \ref{eq:HF_H3}. In each instance quadratic penalty terms favoring $S^{2}=0$ and the correct number of alpha and beta electrons where used with $\lambda$ set to be the $1$-norm of the Hamiltonian. All QUSO problems were then converted to QUBO problems as per Appendix \ref{sec:QUSO2QUBO}. The MaxCut problem was defined using Equation \eqref{eq:maxcut_full}. The penalty terms used to favor particular symmetries (including the ancillary bit) can be found after this equation in the appendix.

All custom SCF and \texttt{\texttt{PySCF}} optimizations were started using the superposition of atomic densities approach \cite{van2006starting}, available in \texttt{PySCF}. At each QUBO/MaxCut SCF iteration an optimization is performed to find the lowest energy single Fock state. To perform each optimization the \texttt{TabuSampler} from \texttt{dwave-tabu} \cite{dwavetabu}, \texttt{SimulatedAnnealingSampler} from \texttt{dwave-neal} \cite{dwaveneal} and the \texttt{CplexOptimizer} in \texttt{qiskit\_optimization} \cite{qiskitopt, qiskit2024} were utilized. 

Each MaxCut problem was solved using the Goemans Williamson (GW) algorithm \cite{goemans1994879, goemans1995improved}. In particular, the SDP problem was solved using \texttt{CVXPY} \cite{cvxpy16} and the Cholesky factorization followed by hyperplane rounding was performed in \texttt{NumPy} \cite{harris2020array}. See Appendix \ref{sec:MaxCut} for further details. As discussed in Appendix \ref{sec:QUSO2MAXCUT}, any solution obtained from  GW with the single ancilla qubit in state $\ket{1}_{w}$ had all the spins flipped (to map the eigenstate to another eigenstate with the same eigenvalue, but in the sector defined by the original Hamiltonian).

For the implementation of the SCF using Algorithm \ref{alg:cap1}, the MO basis ($\vec{\kappa}$) was updated using the orbital gradient \cite[eq 10.8.18]{helgaker2013molecular} and Hessian \cite[eq 10.8.53]{helgaker2013molecular}, each determined using the generalized Fock matrix \cite[eq 10.8.23]{helgaker2013molecular}. This was achieved using the Newton conjugate gradient (CG) method \cite{wright2006numerical} implemented in \texttt{SciPy} \cite{virtanen2020scipy}. Note that we restricted $\vec{\kappa}$ to be real, and thus $U(\vec{\kappa})$ were restricted to be real orthogonal matrices. 

Each RHF calculation performed in \texttt{PySCF} was done using the default RHF class. The second-order (SO) RHF calculation was performed using the newton method in \texttt{PySCF}. An internal stability analysis was also performed on the SO-SCF calculations to verify convergence. If the optimization did not converge, then the SO-RHF calculation was restarted from the  orbitals output by the stability analysis.

Hartree-Fock calculations for \ce{OH-} were also carried out using \texttt{Psi4} \cite{smith2020psi4}, \texttt{Gaussian} \cite{gaussian16}, and \texttt{Orca} \cite{ORCA5}.  The Supplementary Information (SI) provides the input and output files for the \texttt{Gaussian} and \texttt{Orca} studies. Additionally, the output files for \texttt{PySCF} and \texttt{Psi4} are also provided there.

\begin{figure*}
     \centering
     \begin{subfigure}[b]{0.95\textwidth}
         \centering
        \includegraphics[width=0.75\linewidth]{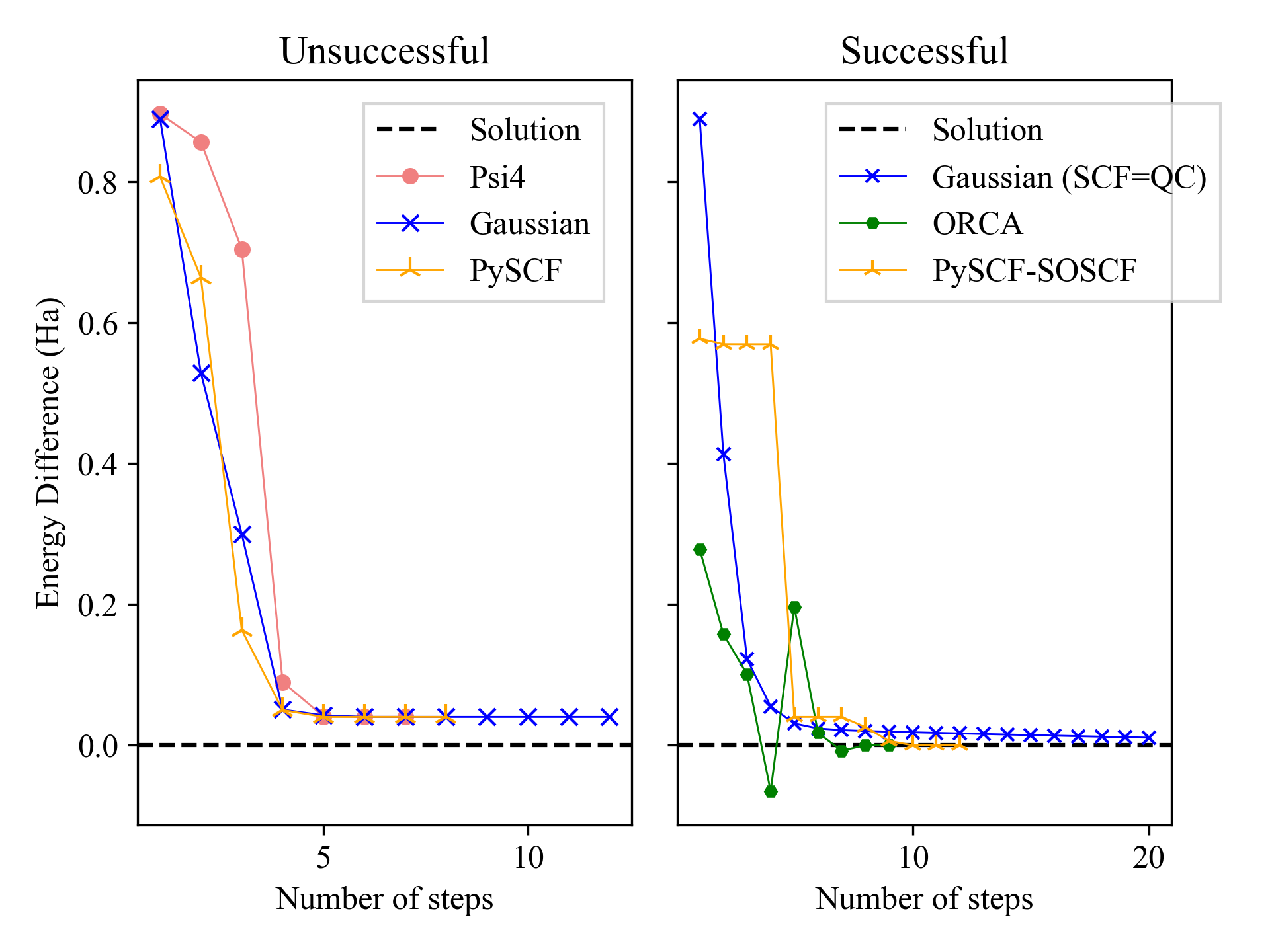}
        \caption{Alternate approaches}
         \label{fig:OH_classical}
     \end{subfigure}
     \begin{subfigure}[b]{0.95\textwidth}
         \centering
         \includegraphics[width=0.75\linewidth]{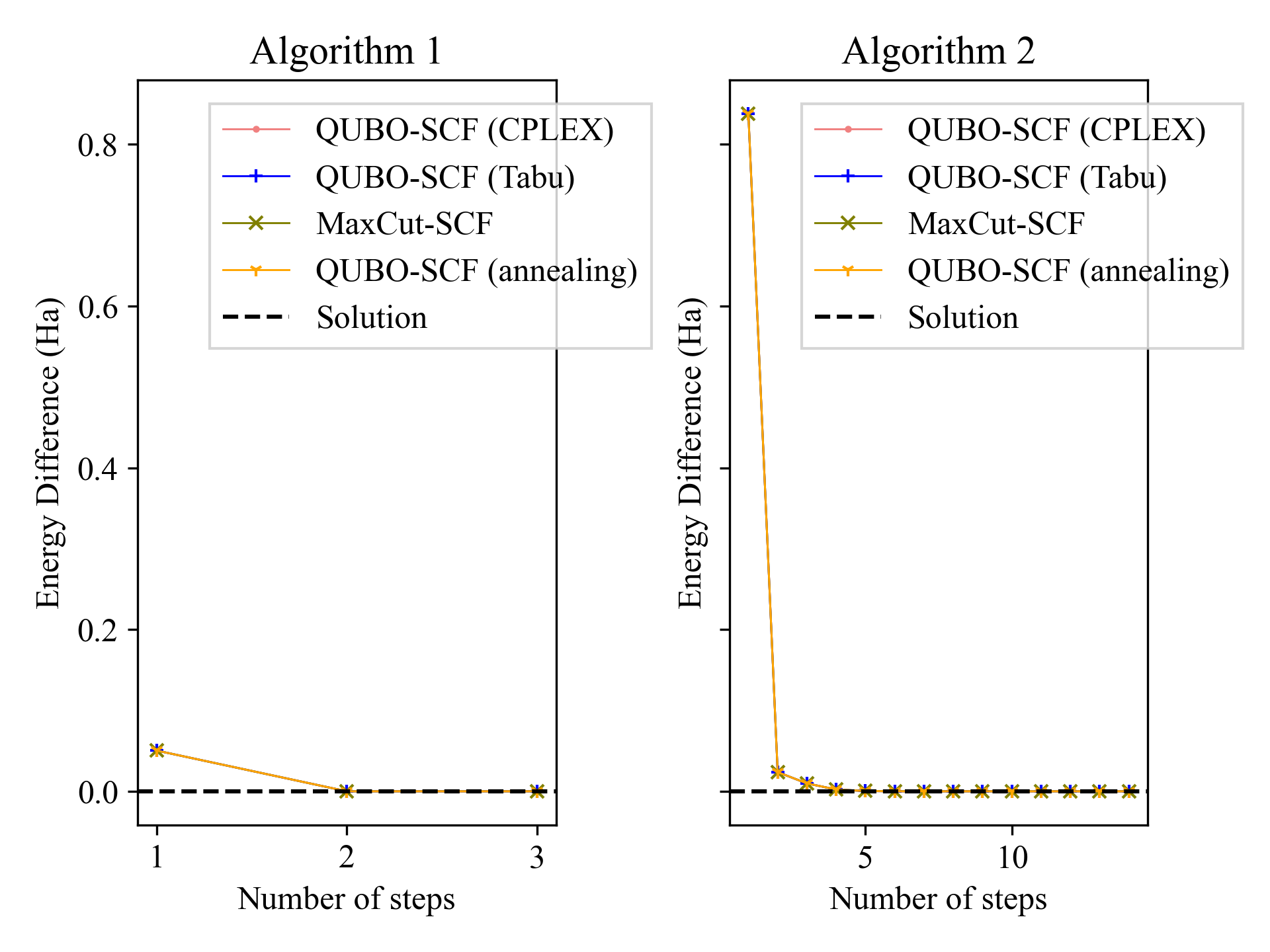}
         \caption{Our work}
         \label{fig:QUBO_SCF_OH}
     \end{subfigure}
        \caption{Converge of different SCF optimization for \ce{OH-} described in the 6-31G basis set (22 qubits/spin-orbitals) with a bond length of $3$ {\AA}. (a) Restricted Hartree-Fock (RHF) performed in conventional chemistry software. (b) The convergence of MaxCut-SCF  and QUBO-SCF solved using different classical optimizers (here annealing is simulated). Figure \ref{fig:algorithms} outlines the steps of Algorithm \ref{alg:cap1}/\ref{alg:cap2}. The raw data for these results may be found in Appendix \ref{sec:OH_DATA}.}
        \label{fig:all_SCF}
\end{figure*}

Note all numerical results presented in this section were obtained using conventional computers; no computations were carried out on quantum hardware.

\begin{figure*}[ht]
    \centering
    \begin{subfigure}[b]{0.45\textwidth}
\includegraphics[width=0.95\linewidth]{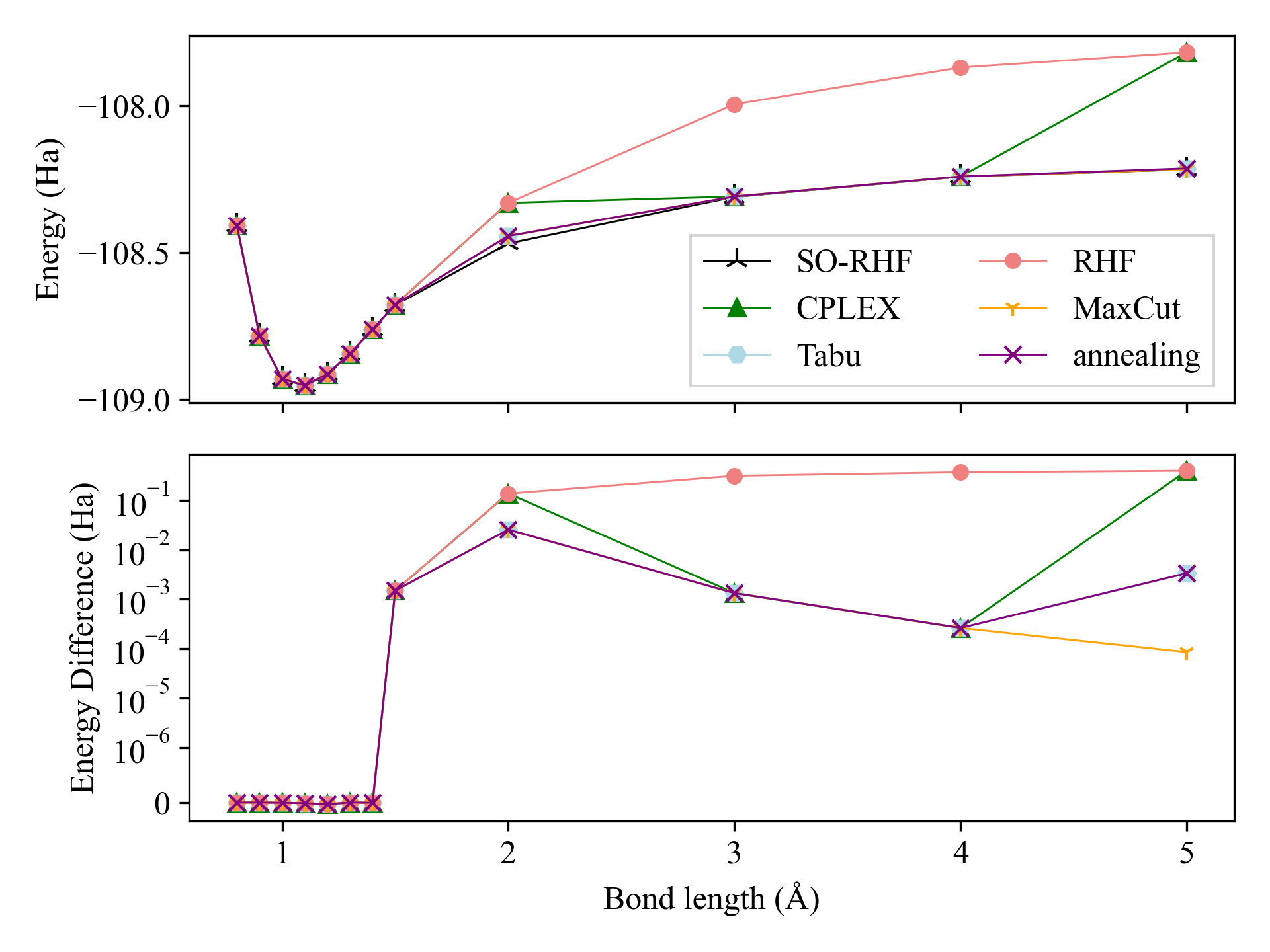}
\caption{cc-pVDZ (56 qubits)} \label{fig:ccpvdz}
\end{subfigure}
    \hspace{0.05\textwidth}
    \begin{subfigure}[b]{0.45\textwidth}
    \includegraphics[width=0.95\linewidth]{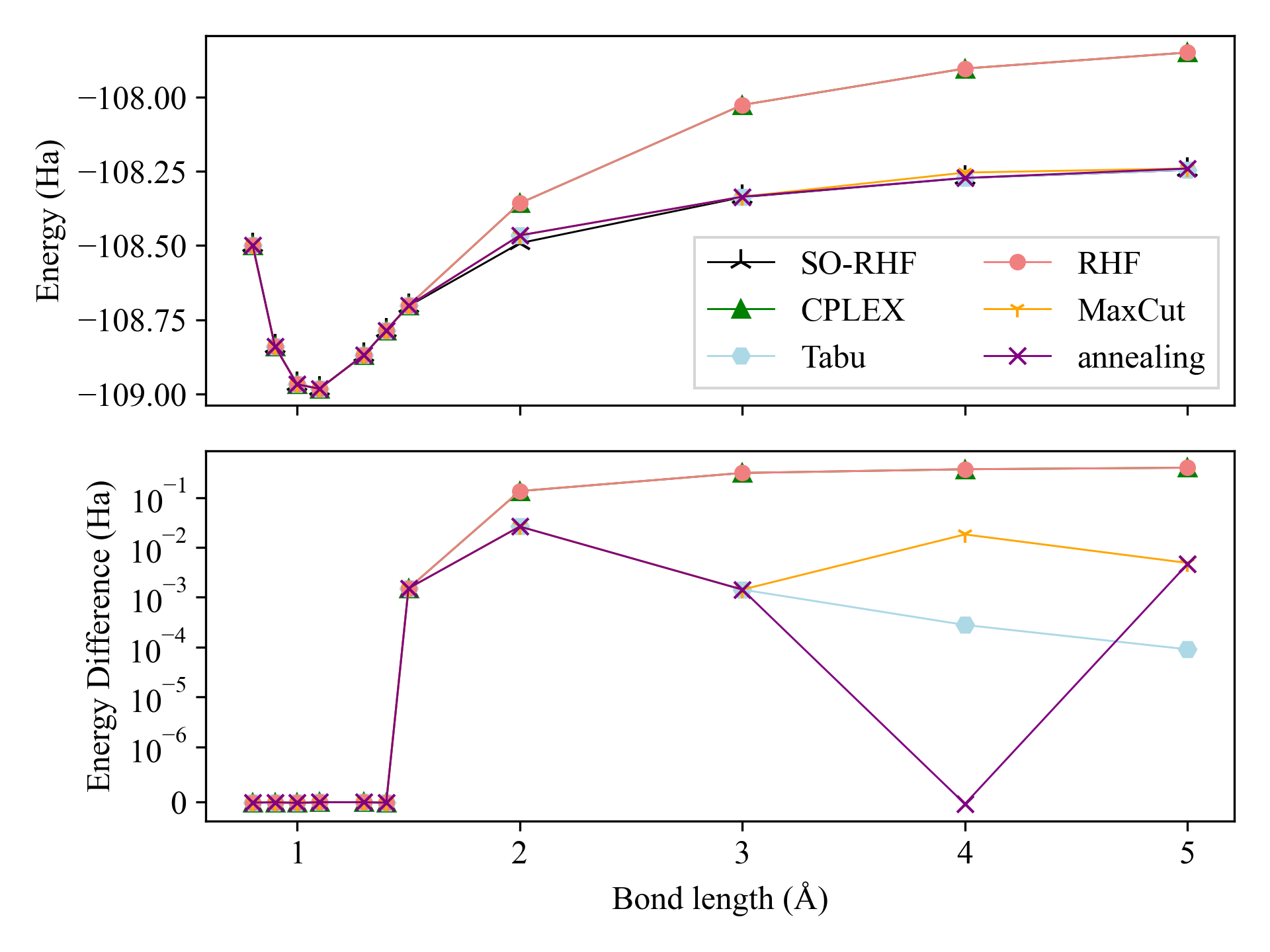}
\caption{cc-pVTZ (120 qubits)} \label{fig:ccpvtz}
    \end{subfigure}
    \hspace{0.05\textwidth}
    \begin{subfigure}[b]{0.45\textwidth}
    \includegraphics[width=0.95\linewidth]{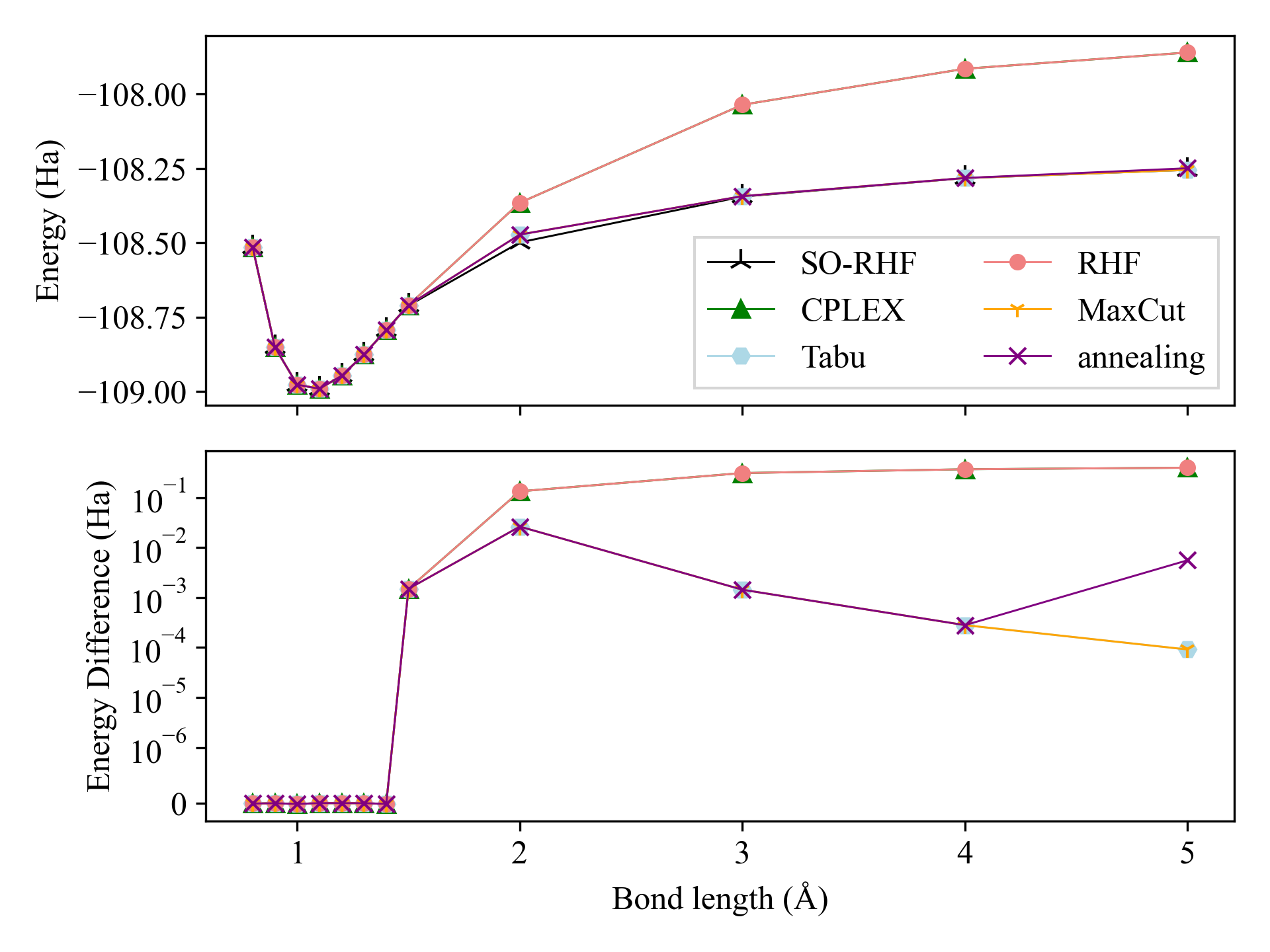}
\caption{cc-pVQZ (220 qubits)} \label{fig:ccpvqz}
    \end{subfigure}
    \caption{SCF potential energy surface for the dissociation of \ce{N2} studied in the (a) cc-pVDZ (56 qubits),(b) cc-pVTZ (120 qubits) and (c) cc-pVQZ (220 qubits) basis sets. The RHF and SO-RHF results are obtained from a calculation performed in \texttt{PySCF}. The internal stability was tracked to ensure the SO-SCF results converged to the true Hartree-Fock ground state. This energy was used as the reference to calculate the energy differences shown in the bottom plots. The other results were obtained from Algorithm $2$ (Figure \ref{fig:alg_overview2}). The raw data for these results may be found in Appendix \ref{sec:N2_DATA}. }
    \label{fig:N2_results}
\end{figure*}

\subsection{Hydroxide Anion} \label{sec:OH}

In Figure \ref{fig:all_SCF}, we study the convergence of different SCF methods compared to QUBO/MaxCut-SCF. First we grouped the SCF results, performed using conventional chemistry codes, that converged to a local minima on the left-hand side of Figure \ref{fig:OH_classical}. These results showed internal instabilities. On the right-hand side of this plot, we show the different conventional SCF approaches that were able to converge to the true solution. Each of these methods required second-order SCF methods to converge to the true Hartree-Fock ground state. 

In comparison,  Figure \ref{fig:QUBO_SCF_OH} shows the convergence of MaxCut/QUBO SCF (i.e. Equation \eqref{eq:HF_vp_QUSO}) implemented by Algorithm \ref{alg:cap1} (Figure \ref{fig:alg_overview}) and Algorithm \ref{alg:cap2} (Figure \ref{fig:alg_overview2}). We observe that all instances converged to the ground state. This result indicates the QUBO- and MaxCut-SCF approaches are more resistant to internal instabilities, which is expected given the optimization over Slater determinants. Furthermore, each method was able to converge to the true ground state within four optimization steps, faster than all approaches in Figure \ref{fig:OH_classical}.

In Figure \ref{fig:OH_classical}, we note that the \texttt{orca} optimization went below the Hartree-Fock energy before the result converged. We believe this is due to the software modifying the underlying problem, such as using level shifting and Direct Inversion in the Iterative Subspace (DIIS), to aid convergence. See the \texttt{orca} raw output for further information.

\begin{figure*}[ht]
    \centering
    \begin{subfigure}[b]{0.45\textwidth}
\includegraphics[width=0.95\linewidth]{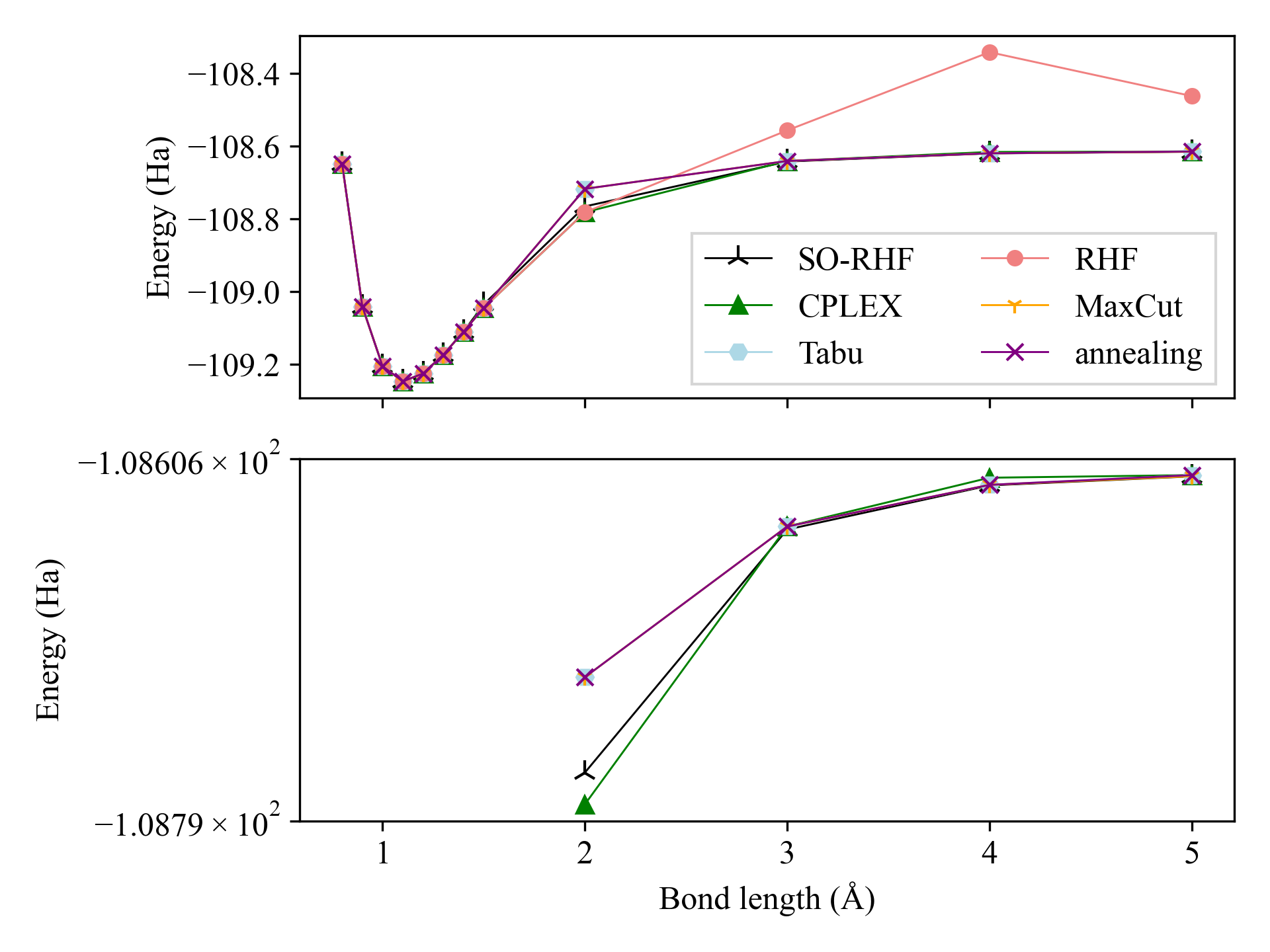}
\caption{cc-pVDZ (56 qubits)} \label{fig:ccpvdz-CISD}
\end{subfigure}
    \hspace{0.05\textwidth}
    \begin{subfigure}[b]{0.45\textwidth}
    \includegraphics[width=0.95\linewidth]{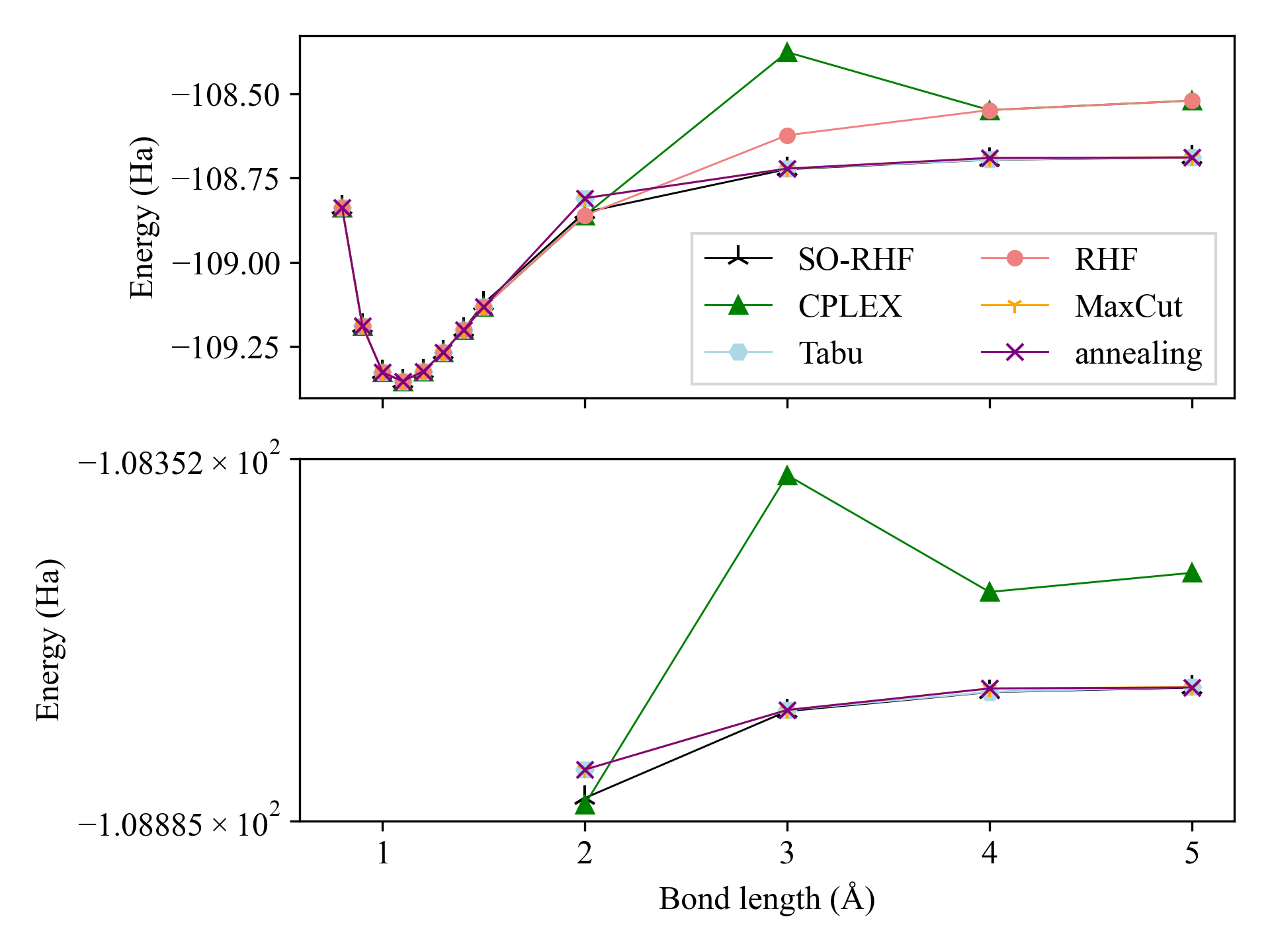}
\caption{cc-pVTZ (120 qubits)} \label{fig:ccpvtz-CISD}
    \end{subfigure}
    \hspace{0.05\textwidth}
    \begin{subfigure}[b]{0.45\textwidth}
    \includegraphics[width=0.95\linewidth]{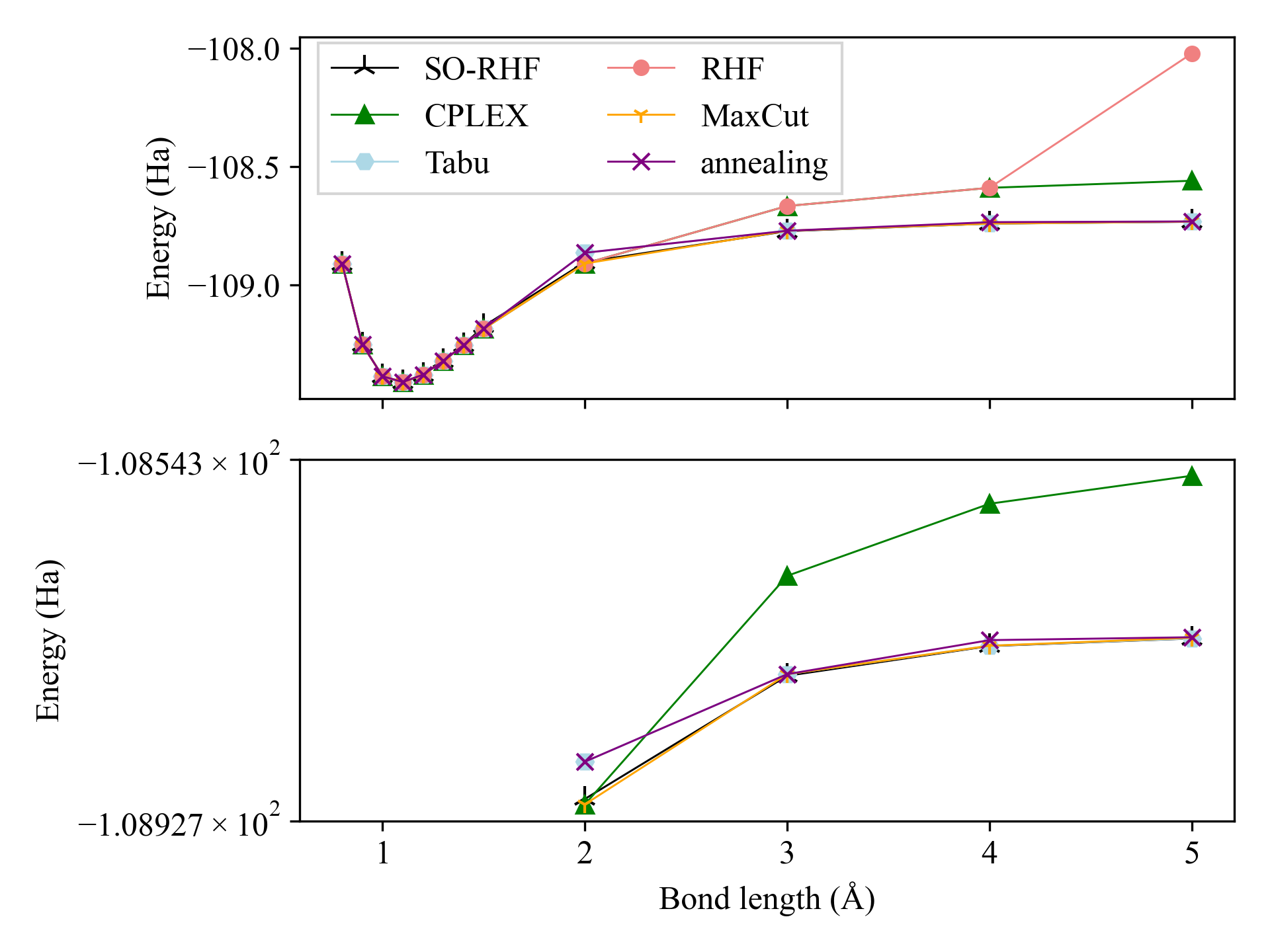}
\caption{cc-pVQZ (220 qubits)} \label{fig:ccpvqz-CISD}
    \end{subfigure}
    \caption{CISD potential energy surface for the dissociation of \ce{N2} studied in the (a) cc-pVDZ (56 qubits),(b) cc-pVTZ (120 qubits) and (c) cc-pVQZ (220 qubits) basis sets. The molecular orbitals utilized for each calculation were those converged in the previous SCF calculation summarised in Figure \ref{fig:N2_results}}
    \label{fig:N2_results-CISD}
\end{figure*}

\subsection{Molecular Nitrogen} \label{sec:N2}

We study the dissociation of molecular Nitrogen \ce{N2} in the cc-pVDZ, cc-pVTZ and  cc-pVQZ basis sets using QUBO-SCF, MaxCut-SCF (Algorithm \ref{alg:cap2}) and compare the results against conventional restricted Hartree-Fock (RHF) calculations performed in \texttt{PySCF}. Two \texttt{PySCF} calculations were performed: the first using the default RHF setup, and the other a second order SCF calculation modified to also eliminate any internal instabilities. Figure \ref{fig:N2_results} shows the results. 

The results indicate that QUBO and MaxCut SCF perform as well as, or better than, conventional RHF for \ce{N2}. At large bond lengths, the RHF method generally underperforms compared to other techniques. This suggests that conventional RHF calculations exhibit internal instabilities for \ce{N2}, as a lower energy single Fock state exists. This is confirmed by the second-order (SO) RHF calculations.

The second-order RHF calculations, in which internal instabilities were checked and corrected when present, performed significantly better. The results were consistent with those obtained using the new SCF approaches. In essence, the QUBO and MaxCut formulation of the single reference SCF optimization is more resistant to internal instabilities. However, they are still possible as seen by the \texttt{cplex} results, in Figure \ref{fig:ccpvdz}, where sometimes the energies calculated matched the RHF calculation when the internal stability was not checked. Note we did not use internal stability methods to try and further optimize the QUBO/MaxCut results, which could be done in these scenarios.

In Figure~\ref{fig:N2_results-CISD}  we show different CISD calculations performed in the MO orbitals selected by QUBO/MaxCut-SCF (further data is given in Appendix \ref{sec:N2_DATA}). The results show the selected basis can be superior to those obtained by conventional SCF (for example see the first \texttt{cplex} data point in Figure \ref{fig:ccpvdz-CISD}). The reason why this is the case is complicated to determine given the high dimensionality of the problem. However, we propose it could be due to the unoccupied MOs being better optimized in the new SCF approaches. This can happen because different Fock states are considered during the iterative optimization procedure. The intuition for why this could be improving post Hatree-Fock calculations relates to the work of Palmieri \textit{et al.} on the optimization of virtual orbitals \cite{palmieri1994hartree}. This is because refinement of the unoccupied orbitals leaves the Hartree-Fock energy unaltered. 


\section{Conclusion} \label{sec:conclusion}

In this work we derive and showcase the QUBO-SCF and MaxCut-SCF algorithms. Both methods are NP-hard, as each requires solving sequential QUBO or MaxCut problems. However, it is well known that Hartree-Fock and Density Function Theory (DFT) are both NP-complete as well \cite{whitfield2014computational, whitfield2014np}. Our QUBO and MaxCut formulation for SCF is thus in the same complexity class as these two methods, but allows a whole new class of optimizers to be applied to the nonlinear optimization problem. 

Moreover, the MaxCut formulation uniquely provides performance guarantees, specifically ensuring the quality of the single Slater determinant state used at each SCF iteration (defining the $1$-RDM) has an energy that is some fraction of optimal in that basis -- an assurance absent from all other existing SCF methods.

Our numerical results, obtained using classical hardware, show that MaxCut-SCF and QUBO-SCF are both competitive with conventional Hartree-Fock solvers including second-order methods, while being resistant to internal instabilities. However, we only studied two molecular systems and further benchmarking of each method is still required.

We prove that the number of positive edge weights of each MaxCut (SCF) problem scales quadratically with the number of spin-orbitals (or qubits), whereas the number of negative edge weights scales linearly with the number of spin-orbitals. This indicates  the approximation ratio given by Equation \eqref{eq:neg_weights2} is likely to be loose. We leave an open question as to whether it can be tightened and if different SDP techniques can be applied to this particular problem to improve the approximation ratio. This may be possible, as this nonlinear optimization problem has a structure arising from the underlying physical model. 

One may consider the possibility of using alternative fermion-to-qubit mappings, such as the Bravyi-Kitaev or parity transformations \cite{bravyi2002fermionic, seeley2012bravyi}. The underlying goal is to map the diagonal Fermionic problem to a $2$-local diagonal Pauli Hamiltonian. From Equation \eqref{eq:HF_H2}, we observe that the HF Hamiltonian is composed of at most a product of two number operators. Therefore, a number encoding of the problem is likely optimal. If the Bravyi-Kitaev or parity encoding was used, the support of the Pauli operators will increase, making the problem no longer quadratic ($2$-local).

Finally, we outlined four hybrid quantum-classical algorithms (GAS-SCF, QAOA-SCF, QA-SCF, DQI-SCF) that each utilize a different quantum subroutine to find the lowest-energy Slater determinant at each SCF step followed by a classical orbital optimization. The high level idea for each procedure is summarized in Equation \eqref{eq:HF_vp_QUSO_quantum}. We leave a thorough investigation of these hybrid approaches for future work.


\section*{Acknowledgements} \label{sec:acknowledgements}
A.R. and P.J.L. acknowledge support by the NSF STAQ project (PHY-1818914/232580) and by the DARPA ONISQ program under ARO Award Number W911NF2320105. T.W. also acknowledges support from the Engineering and Physical Sciences Research Council (EP/S021582/1) and CBKSciCon Ltd.  P.V.C. is grateful for funding from the European Commission for VECMA
(800925) and EPSRC for SEAVEA (EP/W007711/1).  A.R. and P. J. L. acknowledge Kamil Serafin for valuable discussions.

\bibliographystyle{apsrev4-1.bst}
\bibliography{references.bib}

\onecolumngrid
\appendix
\addcontentsline{toc}{section}{Appendices}
\clearpage
\section*{Appendix}
\section{Change of Basis} \label{sec: BCH_fermionic} 
In this appendix, we derive how Equation \eqref{eq:H_trans} is obtained from similarity transformed ladder operators. Note $\hat{k}$ is defined in Equation \eqref{eq:U} in the main text. First, we write the following:
\begin{equation} \label{eq:comm1}
    \begin{aligned} 
        [ \hat{k}, a_{r}^{\dagger} ] &= \sum_{pq} k_{pq} [a_{p}^{\dagger} a_{q} , a_{r}^{\dagger}] \\
         &= \sum_{pq} k_{pq} \big( a_{p}^{\dagger} \underbrace{[ a_{q} , a_{r}^{\dagger} ]}_{=\delta_{qr}} + \underbrace{[ a_{p}^{\dagger}  , a_{r}^{\dagger} ]}_{=0} a_{q} \big) \\
         &= \sum_{pq} k_{pq}  a_{p}^{\dagger} \delta_{qr} \\
         &= \sum_{p} k_{pr}  a_{p}^{\dagger} \equiv k_{pr}  a_{p}^{\dagger} 
    \end{aligned}
\end{equation}
and
\begin{equation} \label{eq:comm2}
    \begin{aligned} 
        [ \hat{k}, a_{r} ] &= \sum_{pq} k_{pq} [a_{p}^{\dagger} a_{q} , a_{r}] \\
         &= \sum_{pq} k_{pq} \big( a_{p}^{\dagger} \underbrace{[ a_{q} , a_{r} ]}_{=0} + \underbrace{[ a_{p}^{\dagger}  , a_{r} ]}_{=-[a_{r}, a_{p}^{\dagger}] = -\delta_{pr}} a_{q} \big) \\
         &= \sum_{pq} - k_{pq}  a_{q} \delta_{pr} \\
         &= \sum_{q} -k_{rq}  a_{q} \\
         &= \sum_{q} k_{qr}^{*}  a_{q} \equiv k_{qr}^{*}  a_{q}, 
    \end{aligned}
\end{equation}
where in the bottom line of Equation \eqref{eq:comm2} we have used the fact that $k_{pq} = -k_{qp}^{*}$ for antihermitian matrices. The final term in both equations uses Einstein summation notation.

For $M$ fermionic modes, the coefficients of $\hat{k}$  can be written inside an $M\times M$ matrix $K$, where $M$ is the number of fermionic modes. The subscript $p,q$ denote the $p,q$-th matrix element, which we write using the notation: $k_{pq} = (K)_{pq}$.

Next, we define the iterated commutator as:

\begin{equation} \label{eq:commut}
    \begin{aligned} 
        [(A)^{n}, B] = \underbrace{[A, \hdots [A, [A}_{n \text{ times}}, B]] \hdots ], \: \: \: \: [(A)^{0}, B] = B.
    \end{aligned}
\end{equation}
Using the results of Equations \eqref{eq:comm1} and \eqref{eq:comm2} and the Baker–Campbell–Hausdorff (BCH) formula if we conjugate $a_{p}^{\dagger}$ with $e^{\hat{k}}$ we obtain the following:

\begin{equation} \label{eq:similarity_dagger}
    \begin{aligned}
        e^{\hat{k}}  a_{p}^{\dagger} e^{-\hat{k}} &= \sum_{n=0}^{\infty} \frac{[(\hat{k})^{n}, a_{p}^{\dagger}]}{n!}
        \\ &= a_{p}^{\dagger}  +  [\hat{k}, a_{p}^{\dagger}]  +  \frac{1}{2} [\hat{k}, [\hat{k}, a_{p}^{\dagger}]] + \frac{1}{6} [\hat{k},[\hat{k}, [\hat{k}, a_{p}^{\dagger}]]] + \hdots \\
        &= a_{p}^{\dagger}  +  a_{r}^{\dagger} k_{pr}  +  \frac{1}{2} [\hat{k}, a_{r}^{\dagger} k_{pr}] + \frac{1}{6} [\hat{k},[\hat{k}, a_{r}^{\dagger} k_{pr}]] + \hdots \\
        &= a_{p}^{\dagger}  +  a_{r}^{\dagger} k_{pr}  +  \frac{(k_{pr} k_{rm})}{2} a_{m}^{\dagger} + \frac{[\hat{k},(k_{pr} k_{rm}) a_{m}^{\dagger}]}{6}  + \hdots \\
            &= a_{p}^{\dagger}  +  a_{r}^{\dagger} k_{pr}  +  \frac{(k_{pr} k_{rm})}{2} a_{m}^{\dagger} +\frac{(k_{pr} k_{rm} k_{mn})}{6} a_{n}^{\dagger}  + \hdots \\
        &= a_{p}^{\dagger}  +  a_{r}^{\dagger} k_{pr}  +  \frac{(K^{2})_{pm}}{2} a_{m}^{\dagger} + \frac{(K^{3})_{pn}}{6} a_{n}^{\dagger} + \hdots \\
        &= a_{p}^{\dagger} (I)_{pr}  +  a_{r}^{\dagger} (K)_{pr}  +  \frac{(K^{2})_{pr}}{2} a_{r}^{\dagger} + \frac{(K^{3})_{pr}}{6} a_{r}^{\dagger} + \hdots  = \sum_{s} a_{s}^{\dagger} [e^{K}]_{ps} \\
        &= a_{p}^{\dagger} (I)_{rp} -  a_{r}^{\dagger} (K)_{rp}  +  \frac{(K^{2})_{rp}}{2} a_{r}^{\dagger} - \frac{(K^{3})_{rp}}{6} a_{r}^{\dagger} + \hdots  = \sum_{s} a_{s}^{\dagger} [e^{-K}]_{sp}.
    \end{aligned}
\end{equation}
Note in the final expressions of the last two lines, we introduce a new general index $s$, that can include both $p$ and $r$ indices. To obtain the final line of this equation we take the normal matrix transpose.

We remark that the matrix $K$ has dimension $M \times M$ (not $2^{M} \times 2^{M}$ which is the dimension of the operator $\hat{k}$ if written as a matrix acting on the full Fock space). The overall transformation can therefore be calculated efficiently, because the dimension of this matrix scales linearly with the number of orbitals and thus qubits (and explicitly $e^{-K}$ can therefore be determined on conventional computers).

By the same steps, one can show:

\begin{equation}  \label{eq:similarity_nondagger}
    \begin{aligned}
        e^{\hat{k}}  a_{p} e^{-\hat{k}} 
        &= a_{p}(I)_{rp} -  a_{r}(K)^{*}_{rp}  +  \frac{(K^{2})^*_{rp}}{2} a_{r}- \frac{(K^{3})^*_{rp}}{6} a_{r}+ \hdots  = \sum_{s} a_{s}[e^{-K}]^{*}_{sp}.
    \end{aligned}
\end{equation}
Equations \eqref{eq:similarity_dagger} and  \eqref{eq:similarity_nondagger} are similarity transformed ladder operators.  Any rotation of the underlying basis can therefore be represented by a similarity transformation of each fermionic mode. To see this, we use the results of Equations \eqref{eq:similarity_dagger} and \eqref{eq:similarity_nondagger} and write:

\begin{equation}
    \begin{aligned} \label{eq:integrals2}
        e^{-\hat{k}} H e^{\hat{k}}=  &\sum_{pq}^{M} h_{pq} e^{-\hat{k}} \alpha_{p}^{\dagger} \alpha_{q} e^{\hat{k}} + \frac{1}{2} \sum_{pqrs}^{M} g_{pqrs} e^{-\hat{k}} \alpha_{p}^{\dagger}  \alpha_{r}^{\dagger}  \alpha_{s} \alpha_{q} e^{\hat{k}} \\
        = &\sum_{pq}^{M} h_{pq} e^{-\hat{k}} \alpha_{p}^{\dagger} \bigg( e^{\hat{k}} e^{-\hat{k}} \bigg) \alpha_{q} e^{\hat{k}}  + \frac{1}{2} \sum_{pqrs}^{M} g_{pqrs} e^{-\hat{k}} \alpha_{p}^{\dagger} \bigg( e^{\hat{k}} e^{-\hat{k}} \bigg) \alpha_{r}^{\dagger}  \bigg( e^{\hat{k}} e^{-\hat{k}} \bigg) \alpha_{s} \bigg( e^{\hat{k}} e^{-\hat{k}} \bigg) \alpha_{q} e^{\hat{k}} \\
        = &\sum_{pq}^{M} h_{pq} \bigg(\sum_{m}^{M} a_{m}^{\dagger} [e^{-K}]_{mp} \bigg) \bigg(\sum_{n}^{M} a_{n} [e^{-K}]_{nq}^{*} \bigg) +\\
        &\frac{1}{2} \sum_{pqrs}^{M} g_{pqrs} \bigg(\sum_{m}^{M} a_{m}^{\dagger} [e^{-K}]_{mp} \bigg)  \bigg(\sum_{n}^{M} a_{n}^{\dagger} [e^{-K}]_{nr} \bigg) \bigg(\sum_{\mu}^{M} a_{\mu} [e^{-K}]_{\mu s}^{*} \bigg) \bigg(\sum_{\nu}^{M} a_{\nu} [e^{-K}]^{*}_{\nu q} \bigg)\\
        = &\sum_{mn}^{M} a_{m}^{\dagger}a_{n}  \sum_{pq}^{M} h_{pq} \bigg( [e^{-K}]_{mp} \bigg) \bigg( [e^{-K}]_{nq}^{*} \bigg) +\\
        &\frac{1}{2}  \sum_{mn\mu \nu}^{M} a_{m}^{\dagger} a_{n}^{\dagger}  a_{\mu} a_{\nu} \sum_{pqrs}^{M} g_{pqrs} \bigg( [e^{-K}]_{mp} \bigg)  \bigg(  [e^{-K}]_{nr} \bigg) \bigg(  [e^{-K}]_{\mu s}^{*} \bigg) \bigg( [e^{-K}]_{\nu q}^{*} \bigg)\\
        = &\sum_{mn}^{M} \bar{h}_{mn}  \alpha_{m}^{\dagger} \alpha_{n} + \frac{1}{2} \sum_{mn\mu \nu}^{M} \bar{g}_{m\nu n \mu} \alpha_{m}^{\dagger}  \alpha_{n}^{\dagger}  \alpha_{\mu} \alpha_{\nu}.
    \end{aligned}
\end{equation}

In the second line of Equation \eqref{eq:integrals2}, we utilize the fact that $e^{\hat{k}} e^{-\hat{k}} = I$, allowing insertion of the identity.  Overall, we see that this change-of-basis can be determined by calculating how the integrals/coefficients of the second quantized molecular Hamiltonian change, explicitly:

\begin{subequations}
\begin{equation} \label{eq:integrals_trans2}
  \bar{h}_{mn} = \sum_{pq}^{M} V(\vec{\kappa})_{mp}V(\vec{\kappa})^{*}_{nq}\cdot {h}_{pq}
\end{equation}    
\begin{equation} \label{eq:integrals_trans}
\begin{aligned}
        \bar{g}_{m\nu n \mu} &= \sum_{pqrs}^{M}  V(\vec{\kappa})_{mp}V(\vec{\kappa})_{nr}V(\vec{\kappa})^{*}_{\mu s} V(\vec{\kappa})^{*}_{\nu q} \cdot g_{pqrs} \\
        &=\sum_{pqrs}^{M}  V(\vec{\kappa})_{mp}V(\vec{\kappa})_{nr}  \cdot g_{pqrs}  \cdot [V(\vec{\kappa})^{\dagger}]_{s \mu} [V(\vec{\kappa})^{\dagger}]_{q \nu },
    \end{aligned}
\end{equation}
\end{subequations}
where
\begin{equation} \label{eq:TimSkew}
    \begin{aligned}
        V(\vec{\kappa}) = e^{-K}, \text{  with  } K = \text{skew}(\vec{\kappa}) = \begin{pmatrix}
0 & k_{12} &  k_{13} & \hdots  & k_{1M}\\ 
-k_{21}^{*} & 0  & k_{23} & \hdots & k_{2M}\\ 
-k_{31}^{*} & -k_{32}^{*} & \ddots  &\hdots  & k_{3M}\\ 
\vdots  & \vdots & \vdots & 0 & \vdots\\ 
-k_{M1}^{*} & -k_{M2}^{*}  & -k_{M3}^{*} & \cdots & 0
\end{pmatrix}.
    \end{aligned}
\end{equation}
Note $V(\vec{\kappa})$ is an $M \times M$ matrix, whereas $U(\vec{\kappa})$ (Equation \eqref{eq:U}) is a $2^{M}\times 2^{M}$. This distinction is why the change of basis can be performed efficiently on a conventional computer. 

We remark that $m,n,\mu, \nu$ are dummy variables that are often re-labelled as $p,q,r,s$ or $p',q',r',s'$ in the literature. We opt to keep the labels $m,n,\mu, \nu$ as it is much clearer.  From Equation \eqref{eq:integrals_trans}, it apparent that obtaining $\bar{g}_{m\nu n \mu}$ will have a cost scaling as $\mathcal{O}(M^{5})$ due to the tensor contraction.

Here $h_{pq}$ and $g_{pqrs}$ are the usual one- and two-electron integrals expressed  in orthonormal molecular orbital (MO) basis: $\int \phi_{p}^{*}(\vec{r}) \phi_{q}(\vec{r}) d\vec{r} = \bra{\phi_{p}} \phi_{q} \rangle = \delta_{pq}$ $\forall p,q$. These are defined as:

\begin{equation}
    \begin{aligned} \label{eq:integrals1_2}
        h_{pq} &= (p| h | q) = \int \phi_{p}^{*}(\vec{r}_{1}) \hat{h}(\vec{r}_{1})  \phi_{q}(\vec{r}_{1}) d\vec{r}_{1} \\ 
        g_{pqrs} &= (pq|rs) \\
        &= \int \phi_{p}^{*}(\vec{r}_{1}) \phi_{r}^{*}(\vec{r}_{2})  \frac{1}{|\vec{r}_{1} - \vec{r}_{2}|}  \phi_{q}(\vec{r}_{1}) \phi_{s}(\vec{r}_{2}) d\vec{r}_{1}d\vec{r}_{2}
    \end{aligned}
\end{equation}
where $\hat{h}(\vec{r}_{1}) =  -\frac{\nabla_{\vec{r}_{1}}^{2}}{2} - \sum_{I}^{N_{\text{atoms}}}  \frac{Z_{I}}{| \vec{r}_{1} - \vec{R}_{I} |}$ is the core Hamiltonian for one electron \cite{helgaker2013molecular, szabo2012modern}.  We note each molecular orbital $\{ \phi_{p} \}$ are represented as a linear combinations of the atomic orbital basis functions $\{ \chi_{\mu} \}$:

\begin{equation}  \label{eq:MO_LCAO}
    \begin{aligned}
        \phi_{p}(\vec{r}) = \sum_{\mu}^{M} C_{\mu p} \chi_{\mu p}(\vec{r}).
    \end{aligned}
\end{equation}
$C$ is the molecular orbital coefficient matrix: an $M \times M$ matrix of orbital coefficients, where column $p$ gives atomic orbital coefficients of molecular orbital $\phi_{p}$. The condition for the MOs defining an orthonormal basis is given by:

\begin{equation}  \label{eq:ortho_def}
    \begin{aligned}
        \delta_{pq} = \sum_{\mu \nu} C_{\mu p}^{*} S_{\mu \nu} C_{\nu q},
    \end{aligned}
\end{equation}
where $S$ is the atomic orbital overlap matrix defined as:

\begin{equation}  \label{eq:overlap}
    \begin{aligned}
        S_{\mu \nu} = \int d\vec{r} \chi_{\mu}^{*}(\vec{r}) \chi_{\nu}(\vec{r}).
    \end{aligned}
\end{equation}

\section{Molecular Symmetries} \label{sec:mol_sym}

In this section, we focus on the common state optimization problem in Equation \eqref{eq:HF_qUso} and Equation \eqref{eq:traditional}. We use $R$ as a placeholder to represent both $\vec{k}$ and $C$. We can enforce symmetries, on both Algorithm \ref{alg:cap1} and Algorithm \ref{alg:cap2}, by constraining $\mathcal{D}$ (the search space):

\begin{equation} \label{eq:HF_qCso}
    \begin{aligned}
    E_{QCSO} &= \min_R   \bigg[ \min_{\ket{b} \in \mathcal{S} \subset \mathcal{D}} \big[ \bra{b} H_{D}(R) \ket{b} \big] \bigg]
    \end{aligned}
\end{equation}
where $\mathcal{S}$ represents the set of single computational basis that respects the symmetry of the underlying problem - for example each element in the set has the correct number of alpha and beta electrons. This is now a quadratic constrained spin optimization (QCSO) problem. A simple example of what $\mathcal{S}$ could look like is the set of all Fock states with the correct number of alpha and beta electrons ($N_{\alpha}$, $N_{\beta}$). The size of such a set for $M/2$-spatial orbitals is $\binom{M/2}{N_{\alpha}} \binom{M/2}{N_{\beta}}$, which has many fewer elements than all $2^{M}$ Fock states (the set $\mathcal{D}$). In conventional language, optimizing over the set $\mathcal{S}$ constrains the optimization to a specific Fock sector. Further constraints can also be imposed to select single Fock states that also respect point group and spin symmetries.

Rather than constraining the search space, penalty terms can be added to penalize elements outside the set $\mathcal{S}$ (i.e., penalize $\mathcal{A} = \mathcal{D} - \mathcal{S}$). This approach favors specific solutions without limiting the search space. In the next subsections, we will see how this reformulation enables the use of advanced solvers.

\subsection{Penalty Terms} 
\label{sec:penalty}
We rewrite the QUSO optimization problem with penalty terms as:
\begin{equation} \label{eq:HF_vp_QUSO_app}
    \begin{aligned}
    E_{QUSO}^{V_{P}} &= \underset{R}{\mathrm{min}}  \bigg[ \underset{\ket{b} \in \mathcal{D}}{\mathrm{min}} \big[ \bra{b} H_{D}(R) + \lambda V_{P} \ket{b} \big] \bigg].
    \end{aligned}
\end{equation}
Here $V_{P}$ penalizes any Slater determinant not in the set $\mathcal{S}$ by $\lambda$, which takes a large positive value.  This provides a way to favour certain molecular symmetries during the QUSO optimization. 

For general quadratic unconstrained spin optimization problems equality constraints can be formulated as:
\begin{equation} \label{eq:spin_constraints)}
    \begin{aligned}
     \sum_{\mathcal{K} \in \mathcal{\chi}} \prod_{i \in \mathcal{K}}  \bigg( \frac{I - Z_{i}}{2} \bigg) = \sum_{\mathcal{K} \in \mathcal{S}} \bar{z}_{\mathcal{K}}= \Omega
    \end{aligned}
\end{equation}
where $\mathcal{\chi}$ represents a set of lists of indices, $\mathcal{K}$ is a particular list of spin orbital indices in $\mathcal{\chi}$ and $\Omega \in \mathbb{R}$ is a constant. In order to impose such a constraint on a particular optimization problem, Equation \eqref{eq:spin_constraints)} can be recast into a penalty term: 
\begin{equation} \label{eq:PenQuso2)}
    \begin{aligned}
    V_{P}(\lambda) &= \lambda \Bigg[ \sum_{\mathcal{K} \in \mathcal{\chi}}  \prod_{i \in \mathcal{K}} \bigg( \frac{I - Z_{i}}{2} \bigg) - \Omega \Bigg]^{2} \\
    &= \lambda \Bigg[ \sum_{\mathcal{K} \in \mathcal{\chi}}  \bar{z}_{\mathcal{K}} - \Omega \Bigg]^{2}
    \end{aligned}
\end{equation}
where $\lambda \gg 0$ is a large number so that the constraint is satisfied. We remark here that in order for the problem to stay quadratic, each $\bar{z}_{\mathcal{K}}$ term in Equation \eqref{eq:PenQuso2)}  can only act nontrivally on a single spin index (i.e. each $\mathcal{K}$ contains at most a single index). This methodology relates to the Lagrangian approach proposed in \cite[section 3.5]{mcclean2016theory}.

In the following Subsections we consider specific molecular symmetries. Our discussion explicitly assumes the Jordan-Wigner transform has been applied to the Fermionic Hamiltonian (Equation \eqref{eq:HF_H2}).

\subsubsection{Number symmetry}

In quantum chemistry, atom(s) and molecules have a specified number of electrons. Therefore, when performing an electronic structure calculation it is necessary to find a solution with the correct number of alpha and beta electrons - i.e. a solution  in the correct Fock sector. 

Such Slater determinants can be favoured in the QUSO optimization by using the following two penalty terms:

\begin{subequations}
\begin{equation} \label{eq:PenQuso_alpha}
    V_{P}^{\alpha}(\lambda) = \lambda \Bigg( \sum_{i \in \mathcal{\chi}_{\alpha}} \bigg( \frac{I - Z_{i}}{2} \bigg) - N_{\alpha}  \Bigg)^{2}
\end{equation}    
\begin{equation} \label{eq:PenQuso_beta}
    V_{P}^{\beta}(\lambda) = \lambda \Bigg( \sum_{i \in \mathcal{\chi}_{\beta}} \bigg( \frac{I - Z_{i}}{2} \bigg) - N_{\beta} \Bigg)^{2}
\end{equation}
\end{subequations}
where $N_{\alpha}$ and $N_{\beta}$ are the number of alpha and beta electrons respectively. Here $\mathcal{\chi}_{\alpha}$ and $\mathcal{\chi}_{\beta}$ give sets of the spin up ($\alpha$) and spin down ($\beta$) MO indices. Each penalty term is added to the objective function (Equation \eqref{eq:QUB)})  As these are both quadratic terms, the overall optimization problem remains a QUSO instance.

\subsubsection{Total spin quantum number}
In electronic structure calculations it is also desired to have a solution in the correct spin state \cite{chen1994evaluation}.  For example, single-triplet instabilities often occur during covalent bond breaking processes \cite{yamada2015singlet}, and thus being able to favor a specific $\langle S^{2} \rangle$ can be useful in such scenarios. Here $S$ is the spin angular momentum operator. For the construction of these operators see \cite[Section 3]{yen2019exact}. The eigenvalues of $S^{2}$ are $s(s+1)$ (in units of $\hbar$), where $s$ is called spin quantum number or just spin. This relates to the spin multiplicity which is given as $2s + 1$ (the different values of the spin projection quantum number $m_{s}$ for a given $s$). States with particular multiplicities are often denoted as singlet ($s=0$), doublet ($s=1/2$), triplet ($s=1$)...etc states. 

The $S^{2}$ operator is written as  \cite[Section 2.5.1]{szabo2012modern}\cite{yen2019exact}:

\begin{equation} \label{eq:S2_op}
    \begin{aligned}
S^{2} = S_{z} + S_{z}^{2} + S_{-} S_{+}.
    \end{aligned}
\end{equation}
However, as we are limited to only single Fock states only the diagonal components of this operator are necessary. In Appendix \ref{sec:app_A} we take a closer look at what this means in both the restricted and restricted settings.

For closed shell \footnote{The number of spin-up and -down electrons are equal.} restricted systems, we note that the value of $S^{2}$ is determined simply by the number of unpaired $\beta$ electrons (see Appendix \ref{sec:app_A}). Therefore, to favor $\langle S^{2}\rangle =0 $ it is possible to favor doubly occupied/empty spatial orbitals. This can be done as follows:

\begin{equation} \label{eq:PenQusoSpin}
    \begin{aligned}
    V_{P}^{S^{2}}(\lambda) &= \lambda \Bigg[ N_{\beta} - \sum_{k=0}^{M/2 -1}  \bigg( \frac{I + Z_{2k} Z_{2k+1}}{2} \bigg) - \underbrace{S^{2}}_{\text{desired value}} \Bigg]^{2}.
    \end{aligned}
\end{equation}
Here  $M/2$ is the number of spatial orbitals and each $2k$, $2k+1$ pair are spin orbitals  that share the same spatial orbital. The $(I + Z_{2k} Z_{2k+1}) /2$ is simply counting the number of spatial orbitals that are doubly occupied/unoccupied exactly as required for closed shell restricted systems. Due to the squaring operation this operator will contain quartic terms and thus cannot just be added to the optimization problem as the overall problem would then not be  quadratic. Instead, quadratization techniques must first be used to convert the $V_{P}^{S^{2}}(\lambda)$ to a quadratic penalty and then added to the optimization problem. For the same reason if the term in Equation \eqref{eq:S2_op} was used this would also  be required.  In Appendix \ref{sec:quadratization}, we discuss how this can be achieved. 

Fortunately, for closed shell systems if $S^{2}=0$  is required the penalty term above can be modified. Instead the following can be used:

\begin{equation} \label{eq:PenQusoSpin0}
    \begin{aligned}
    V_{P}^{S^{2}}(\lambda) &= \lambda \Bigg[ N_{\beta} - \sum_{k=0}^{M/2 -1}  \bigg( \frac{I + Z_{2k} Z_{2k+1}}{2} \bigg)  \Bigg].
    \end{aligned}
\end{equation}
This operator is always non-negative for closed shell systems (and thus doesn't requiring squaring), and only equals zero for spin paired sites thus favoring Slater determinants with spatial orbitals doubly occupied or empty - which is required for $\langle S^{2} \rangle = 0$, as discussed in Appendix \ref{sec:app_A}. Note this penalty is quadratic, so can be added to the QUSO problem without requiring quadratization techniques. We note this term works best if the $N_{\alpha}$ and $N_{\beta}$  penalty operators are included to favor closed shell states. Otherwise, the more complicated $S^{2}$ operator defined in Equation \eqref{eq:S2_op} can be utilized (still quadratic under the Jordan-Wigner transform).

Generalization of these penalty operators to the unrestricted setting can be done, but would require the MO overlap matrix of the alpha and beta molecular orbitals, which is also discussed in the Appendix \ref{sec:app_A}.

In the restricted formalism all single Slater determinant solutions are eigenfunctions of $S^{2}$, but enforcing a particular value \textit{a priori} requires restricting the search space or penalizing solutions (eigenstates) in the wrong symmetry sector. However, since all solutions QUSO solutions will be eigenstates of $\langle S^{2} \rangle$, an optimization can be run without penalizing for a specific $\langle S^{2} \rangle$ value and the solution checked to see if $\langle S^{2} \rangle$ has the desired value. If the result has the wrong symmetry, it can simply be re-run with the search space reduced to only include terms with the correct symmetry or the penalization term (Equation \eqref{eq:PenQusoSpin}) added.

Finally, Yen \textit{et al.} comment in \cite{yen2019exact} on how the $S_{z}$ symmetry operator can be used to favor non-singlet eigenstates. This idea can be directly applied here.

\section{Total Spin Quantum Number} \label{sec:app_A}

For single-determinant wavefunctions, such as restricted Hartree-Fock (RHF), restricted open-shell Hartree-Fock (ROHF) and unrestricted Hartree-Fock (UHF), the total spin quantum number is obtained as \cite[eq 4]{andrews1991spin}:

\begin{equation}
    \langle S^{2} \rangle = \bigg( \frac{N_{\alpha} - N_{\beta}}{2} \bigg)^{2} + \bigg( \frac{N_{\alpha} - N_{\beta}}{2} \bigg) + N_{\beta} - \sum_{j \in O_{\alpha}} \sum_{k \in O_{\beta}} |\bra{\psi_{j}^{\alpha}} \psi_{k}^{\beta} \rangle |^{2}
\end{equation}
where $O_{\alpha}$ and $O_{\beta}$ are sets containing occupied spatial orbital indices for alpha and beta electrons. As the spatial orbitals in RHF/ROHF are fixed to be the same for pairs of spin orbitals, we find that $\bra{\psi_{j}^{\alpha}} \psi_{k}^{\beta} \rangle = \delta_{jk}$, because in general for an orthonormal molecular orbital basis:

\begin{equation}  \label{eq:ortho_def}
    \begin{aligned}
        \bra{\psi_{p}} \psi_{q} \rangle  = \sum_{\mu \nu} C_{\mu p}^{*} S_{\mu \nu} C_{\nu q} = \delta_{pq} .
    \end{aligned}
\end{equation}
Here $S_{\mu \nu}$ is the atomic orbital overlap matrix defined as:

\begin{equation}  \label{eq:overlap}
    \begin{aligned}
        S_{\mu \nu} = \int d\vec{r} \chi_{\mu}^{*}(\vec{r}) \chi_{\nu}(\vec{r}).
    \end{aligned}
\end{equation}

Overall, we can write the total spin squared value for ROHF as:

\begin{equation} \label{eq: ROHF_Eq}
\begin{aligned}
    \langle S_{ROHF}^{2} \rangle &= \bigg( \frac{N_{\alpha} - N_{\beta}}{2} \bigg)^{2}  + \bigg( \frac{N_{\alpha} - N_{\beta}}{2} \bigg) + N_{\beta} - \sum_{j \in O_{\alpha}} \sum_{k \in O_{\beta}} \delta_{jk} \\
    &= \bigg( \frac{N_{\alpha} - N_{\beta}}{2} \bigg)^{2}  + \bigg( \frac{N_{\alpha} - N_{\beta}}{2} \bigg) + N_{\beta} - N_{\text{paired}} \\
&= \bigg( \frac{N_{\alpha} - N_{\beta}}{2} \bigg)^{2}  + \bigg( \frac{N_{\alpha} - N_{\beta}}{2} \bigg) + N_{\text{unpaired }\beta} 
\end{aligned}
\end{equation}

The way ROHF is usually performed,is doubly occupied spatial orbitals are used as far as possible and then the remaining  electrons are placed in spatial orbitals such that they are singly occupied. From Equation \eqref{eq: ROHF_Eq}, it is easy to see why if $N_{\alpha} = N_{\beta}$ and thus $O_{\alpha} = O_{\beta}$ (due to doubly occupied spatial orbitals condition), then $\langle S_{ROHF}^{2} \rangle =0$ and is thus equivalent to RHF. 

However, if we consider the case of $N_{\alpha} = N_{\beta}$ but don't require all spatial orbitals to be doubly occupied as much as possible ($O_{\alpha} \neq O_{\beta}$)  then we observe that $\langle S_{ROHF}^{2} \rangle  \neq 0$ and instead its value is determined by the number indices that differ between the sets $O_{\alpha}, O_{\beta}$. Overall, the final two terms in Equation \eqref{eq: ROHF_Eq} will not cancel out leading to nonzero values of $\langle S_{ROHF}^{2} \rangle$. Thus for a closed shell system, if the condition that all spatial orbitals must be double occupied is relaxed then it is possible for $\langle S_{ROHF}^{2} \rangle$ to differ from $\langle S_{RHF}^{2} \rangle$, while in both cases all solutions will remain a true eigenstate of $S^{2}$ (unlike UHF).

To enforce $S^{2}$ in the unrestricted setting, the MO overlap term in Equation \eqref{eq:ortho_def} must be accounted for in any penalization term. It is important to note single Fock states may not be eignstates of $S^{2}$ due to these overlap terms. In such settings, enforcing a particular eigenvalue of $S^{2}$ is not possible; however, penalization terms can be constructed to favor a particular value without the solutions being true eigenstates, thus limiting the amount of spin contamination.






\section{Quadratizations of pseudo-Boolean functions \label{sec:quadratization}}


In general, the second quantized molecular Hamiltonian, under the Jordan-Wigner transformation, is written as a linear combination of Pauli operators many of which act non-trivially on more than $2$ qubits and also includes off-diagonal terms. The technique of perturbation gadgets can be used to build these as two-body ones \cite{kempe2006complexity, jordan2008perturbative, bravyi2008quantum, babbush2013resource, bausch2020perturbation, cichy2022perturbative}. This is achieved by increasing the dimension of the Hilbert space (by adding ancilla qubits) and `encoding' the target Hamiltonian in the low energy subspace of the `gadget' Hamiltonian . Such a strategy is a common tool in adiabatic quantum computing. However, as we have shown in this work the Hartree-Fock Hamiltonian is already in the correct form to define as a quadatric unconstratined spin optimization problem. 

However, imposing penalty terms (such as Equation \eqref{eq:PenQusoSpin}) can result in a polynomial unconstrained spin optimization problem that is not quadratic. If the problem remains diagonal, we can apply classical quadratization techniques for pseudo-Boolean functions. This is exactly the regime for QUBO-SCF, as solutions are restricted to be single computational basis states and thus only diagonal Hamiltonian terms will have non-zero expectation values.




Rosenberg showed in 1975 that any pseudo-Boolean function admits a quadratization and that such a mapping can be computed in polynomial time from the polynomial expression of $f(\vec{x})$ \cite{rosenberg1975reduction} such that:

\begin{equation}
    f(\vec{x}) = \underset{\vec{y}\in \{0,1\}^{m}}{\mathrm{min}} \big\{ g(\vec{x}, \vec{y})  \big\} \: \: \: \: \forall \: \vec{x} \in \{0,1\}^{n}.
\end{equation}
Here $g(\vec{x}, \vec{y})$ is a quadratization of $f(\vec{x})$ and is a quadratic polynomial depending on $\vec{x}$ and $m$ auxiliary variables $\{ y_{0},y_{1}, \hdots, y_{m-1} \}$. We note the problem of optimizing the quadratic reformulation is still NP-hard \cite{boros2020compact}.

One approach to achieve such a reduction is the substitution approach proposed by Rosenberg that was rediscovered in the context of diagonal quantum Hamiltonians \cite{Biamonte2008}. The way this mapping is implemented is as follows:

\begin{enumerate}
    \item Select an arbitrary cubic term $x_{i}x_{j}x_{k}$
    \item Replace with $x_{i}x_{j}x_{k} \mapsto x_{a}x_{k} + x_{i}x_{j} - 2 x_{i}x_{a} - x_{j}x_{a} + 3x_{a}$, yielding a $(k-1)$-local function. Note $x_{a}$ is an extra ancillary bit.
    \item Repeat steps 1 and 2 (in total $k-2$ times) until the resulting function is $2$-local.
\end{enumerate}




Alternate transformations have been also been given such as by Boros and Hammer in \cite[Section 4.4]{Boros2002}, Buchheim and Rinaldi in \cite{Williams2009} and Elloumi \textit{et al.} in \cite{Elloumi2021}. We note that one advantage of quadratization techniques over perturbation gadgets is that auxiliary qubits (or perturbations) are not required for certain classical quadratization techniques. For example \cite{tanburn2014reducing} uses no ancillary qubits.
\section{Hamiltonian Mappings} \label{sec:MaxCutChem}
In this section we write down the mapping of the diagonal  Hamiltonian (equation  \ref{eq:HF_H2}) from QUSO to QUBO and MaxCut respectively. Note that the dependence of each coefficient on $\vec{\kappa}$ or $C$ is omitted for clarity. We leave the functional dependence of the Hamiltonian on $R$ to indicate this fact.

\subsection{QUSO to QUBO mapping} \label{sec:QUSO2QUBO}
In the main text it is shown that that diagonal  Hamiltonian (equation  \ref{eq:HF_H2}) is mapped to a QUSO under the Jordan Wigner transform:

\begin{equation} 
    \begin{aligned} \label{eq:QUSO_full}
        H_{D}(R) &= \sum_{n}^{M} \bar{h}_{nn} \bigg(\frac{I - Z_{n}}{2}\bigg) +
        \frac{1}{2} \sum_{\substack{m,n \\ m \neq n}}^{M} \frac{\bar{g}_{mmnn}- \bar{g}_{mnnm}}{4} \bigg(I - Z_{n} - Z_{m} + Z_{m}Z_{n} \bigg)\\
        &\mapsto \\
        H_{QUSO}(R) &= \sum_{n}^{M} \bar{h}_{nn} \bigg(\frac{1 - z_{n}}{2}\bigg) +
        \frac{1}{2} \sum_{\substack{m,n \\ m \neq n}}^{M} \frac{\bar{g}_{mmnn}- \bar{g}_{mnnm}}{4} \bigg(1 - z_{n} - z_{m} + z_{m}z_{n} \bigg)
    \end{aligned}
\end{equation}
This problem represents the quadratic unconstrained spin optimization (QUSO) minimization problem. As this problem is optimized over single Fock states, we replace each $Z_{i}$ operator with a spin variable $z_{i} \in \{-1, +1 \}$. The QUSO problem is then optimized over a vector $\vec{z} \in \{+1,-1 \}^{M}$, which sets the expectation value of $H_{QUSO}$. 

This problem can be converted into a QUBO by mapping the spin variables to binary variables via $z_{i} \mapsto (1-2x_{i})$. This results in:

\begin{equation} \label{eq:qubo_full}
    \begin{aligned}
H_{QUBO}(R) &= \sum_{n}^{M} \bar{h}_{nn} \bigg(\frac{1 - (1-2x_{n})}{2}\bigg) +
        \frac{1}{2} \sum_{\substack{m,n \\ m \neq n}}^{M} \frac{\bar{g}_{mmnn}- \bar{g}_{mnnm}}{4} \bigg(1 - 
        (1-2x_{n}) - (1-2x_{m}) +  (1-2x_{n}) (1-2x_{m})\bigg) \\
        &= \sum_{n}^{M} \bar{h}_{nn} \big(x_{n}\big) +
        \frac{1}{2} \sum_{\substack{m,n \\ m \neq n}}^{M} \frac{\bar{g}_{mmnn}- \bar{g}_{mnnm}}{4} \big(4 x_{n}x_{m} \big) \\
        &= \sum_{n}^{M} \bar{h}_{nn} \big(x_{n}\big) +
        \frac{1}{2} \sum_{\substack{m,n \\ m \neq n}}^{M} \bigg( \big[ \bar{g}_{mmnn}- \bar{g}_{mnnm} \big] x_{n}x_{m} \bigg).
    \end{aligned}
\end{equation}
Equation \ref{eq:qubo_full} defines the QUBO minimization problem.

\subsection{QUSO to MaxCut mapping} \label{sec:QUSO2MAXCUT}

In \cite[Section 3]{boros1991max} Boros and Hammer show how a QUBO problem can be mapped onto a Max-Cut problem using a single ancilla/auxilary bit. We use this procedure to inspire a mapping for the QUSO problem.

The MaxCut cost function is defined as follows:

\begin{equation} \label{eq:maxcut_cost}
    \begin{aligned}  
       C_{MaxCut} &=  \sum_{ij}^{M} A_{ij} \bigg( \frac{I-Z_{i}Z_{j}}{2}\bigg) \\
       &= \sum_{ij}^{M} \bigg( \frac{A_{ij} I}{2} +\frac{-A_{ij}Z_{i}Z_{j}}{2}\bigg)
    \end{aligned}
\end{equation}
where $A_{ii}=0 \; \forall i$. The goal of MaxCut is to find the state which maximizes this function. 

To convert the diagonal fermionic Hamiltonian (equation  \ref{eq:HF_H2}):

\begin{subequations}
\begin{equation}  
    \begin{aligned} \label{eq:Ferm1}
        &H_{D}^{\text{ferm}}(R) 
        = \sum_{n}^{M} \bar{h}_{nn}  \alpha_{n}^{\dagger} \alpha_{n} +
        \frac{1}{2} \sum_{\substack{m,n \\ m \neq n}}^{M}  \bigg( [\bar{g}_{mmnn}- \bar{g}_{mnnm}]  \alpha_{m}^{\dagger} \alpha_{m} \alpha_{n}^{\dagger} \alpha_{n}   \bigg) \\
        &\mapsto
    \end{aligned}
\end{equation}
\begin{equation} \label{eq:JW1}
\begin{aligned}
H_{D}^{JW}(R) = \sum_{n}^{M} \bar{h}_{nn} \bigg(\frac{I - Z_{n}}{2}\bigg) +
        \frac{1}{2} \sum_{\substack{m,n \\ m \neq n}}^{M} \frac{\bar{g}_{mmnn}- \bar{g}_{mnnm}}{4} \bigg(I - Z_{n} - Z_{m} + Z_{m}Z_{n} \bigg).
    \end{aligned}
\end{equation}
\end{subequations}
into MaxCut form, we first need to add a single  ancillary fermionic spin at index $\omega$. This should not be an existing spin index. Without loss of generality, let $\omega = 0$:

\begin{subequations}
\begin{equation} \label{eq:Ferm2}
    \begin{aligned}  
        H_{\omega}^{\text{ferm}}(R) 
        =& \sum_{n=1}^{M} \bar{h}_{nn}  \bigg(\alpha_{n}^{\dagger} \alpha_{n} + \alpha_{\omega}^{\dagger} \alpha_{\omega} - 2 \alpha_{n}^{\dagger} \alpha_{n} \alpha_{\omega}^{\dagger} \alpha_{\omega} \bigg) + \\
        &\frac{1}{2} \sum_{\substack{m,n=1 \\ m \neq n}}^{M}  \Bigg( [\bar{g}_{mmnn}- \bar{g}_{mnnm}]  \bigg( \alpha_{\omega}^{\dagger} \alpha_{\omega} + \alpha_{m}^{\dagger} \alpha_{m} \alpha_{n}^{\dagger} \alpha_{n} - \alpha_{n}^{\dagger} \alpha_{n}\alpha_{\omega}^{\dagger} \alpha_{\omega} - \alpha_{m}^{\dagger} \alpha_{m}   \alpha_{\omega}^{\dagger} \alpha_{\omega}\bigg)  \Bigg)\\
        &\mapsto
    \end{aligned}
\end{equation}
\begin{equation} \label{eq:JW2}
    \begin{aligned}  
        H_{\omega}^{\text{JW}}(R) 
        =& \sum_{n=1}^{M} \bar{h}_{nn} \bigg(\frac{I - Z_{n}Z_{\omega}}{2}\bigg) +
        \frac{1}{2} \sum_{\substack{m,n=1 \\ m \neq n}}^{M} \frac{\bar{g}_{mmnn}- \bar{g}_{mnnm}}{4} \bigg(I - Z_{n}Z_{\omega} - Z_{m}Z_{\omega} + Z_{m}Z_{n} \bigg).
    \end{aligned}
\end{equation}
\end{subequations}

Under the Jordan-Wigner transformation, we see that this relates to Equation \eqref{eq:JW2} relates to Equation \eqref{eq:JW1} except that all single Pauli $Z$ operators have been modified to have an additional $Z$ acting on index $\omega$. This means that all operators are two local, as needed by the MaxCut cost function. 

Furthermore, inspecting Equation \eqref{eq:Ferm2}, we note that the spin on index $\omega$ can be in the state $\ket{0}_{\omega}$ or $\ket{1}_{\omega}$. For $\ket{0}_{\omega}$, it is trivial to see that the original Hamiltonian is returned as: $\alpha_{\omega}^{\dagger} \alpha_{\omega} \ket{0} =0$. For the case of $\ket{1}_{\omega}$ we find:

\begin{equation} \label{eq:Ferm3}
    \begin{aligned}  
        H_{\ket{1}_{\omega}}^{\text{ferm}}(R) 
        =& \sum_{n}^{M} \bar{h}_{nn}  \bigg( 1 - \alpha_{n}^{\dagger} \alpha_{n}  \bigg) + \\
        &\frac{1}{2} \sum_{\substack{m,n \\ m \neq n}}^{M}  \Bigg( [\bar{g}_{mmnn}- \bar{g}_{mnnm}]  \bigg( 1 + \alpha_{m}^{\dagger} \alpha_{m} \alpha_{n}^{\dagger} \alpha_{n} - \alpha_{n}^{\dagger} \alpha_{n} - \alpha_{m}^{\dagger} \alpha_{m} \bigg)  \Bigg)
    \end{aligned}
\end{equation}

Interestingly, there is a 1:1 correspondence with the original fermionic Hamiltonian except that for each eigenvalue the corresponding eigenstate is the corresponding eigenstate of the original Hamiltonian with all the spins flipped (the bitwise-not bitstring).

To see this, consider the first term in equation  \ref{eq:Ferm3}. The $\bar{h}_{nn}$ term only contributes if there is no fermion in orbital $n$ as opposed to the original Hamiltonian when a fermion in orbital $n$ is required for this term to contribute (Equation \eqref{eq:Ferm1}). Likewise, for $[\bar{g}_{mmnn}- \bar{g}_{mnnm}]$ to contribute to the energy requires orbitals $n$ and $m$ to both be empty; whereas, in the original Hamiltonian this term only contributes when both orbitals are occupied. 

In summary, the fermionic eigenstates for the original Hamiltonian (equivalent to when the $\omega$ bit is fixed in the $\ket{0}_{\omega}$ state) is matched by an eigenstate with the same eigenvalue where all spins are all flipped (bitwise not) with the $\omega$ bit always in the $\ket{1}_{\omega}$ state. This structure lets us graphically represent the Hamiltonian as:

\begin{equation}
\begin{aligned}
    H_{\omega}^{\text{ferm}}(R) = \begin{pNiceMatrix}[margin]
\Block[draw,fill=blue!15,rounded-corners]{4-4}{}
a & 0 & 0 & 0 & 0 & 0 & 0 & 0 \\
0 & b & 0 & 0 & 0 & 0 & 0 & 0 \\
0 & 0 & \ddots & 0 & 0 & 0 & 0 & 0 \\
0 & 0 & 0 & c & 0 & 0 & 0 & 0 \\
0 & 0 & 0 & 0 &\Block[draw,fill=red!15,rounded-corners]{4-4}{}
                c & 0 & 0 & 0 \\
0 & 0 & 0 & 0 & 0 & b & 0 & 0 \\
0 & 0 & 0 & 0 & 0 & 0 & \ddots & 0 \\
0 & 0 & 0 & 0 & 0 & 0 & 0 & a
\end{pNiceMatrix} = \ket{0}_{\omega}\bra{0}_{\omega} \otimes
&\begin{pNiceMatrix}[margin]
\Block[draw,fill=blue!15,rounded-corners]{4-4}{}
a & 0 & 0 & 0 \\
0 & b& 0 & 0 \\
0 & 0 & \ddots & 0 \\
0 & 0 & 0 & c
\end{pNiceMatrix} 
+ \ket{1}_{\omega}\bra{1}_{\omega} \otimes \begin{pNiceMatrix}[margin]
\Block[draw,fill=red!15,rounded-corners]{4-4}{}
c & 0 & 0 & 0 \\
0 & b& 0 & 0 \\
0 & 0 & \ddots & 0 \\
0 & 0 & 0 & a
\end{pNiceMatrix}
\end{aligned}
\end{equation}
for $\omega=0$. For other $\omega$ indices this will remain true up to permutations of the diagonal matrix elements. This is important for the MaxCut optimization problem, as any result obtained in the bottom right (red) sector can be converted to the state in the upper left (blue) sector by taking the bitwise not of each spin. This is easy to do as when the spin at site $\omega$ is in the state $\ket{1}_{w}$ all the bits can just be flipped. Crucially, this correction makes sure no samples are wasted when approximately solving the problem using randomized methods (Appendix \ref{sec:MaxCut}). 

To determine the graph edges of our problem in the form of the MaxCut problem, we first need to multiply Equation \eqref{eq:JW2} by $-1$ to turn the minimization problem into a maximization:

\begin{equation} \label{eq:JW3}
    \begin{aligned}  
        H_{MaxCut}^{\text{JW}}(R) 
        &= \sum_{n=1}^{M} -\bar{h}_{nn} \bigg(\frac{I - Z_{n}Z_{\omega}}{2}\bigg) +
         \sum_{\substack{m,n=1 \\ m \neq n}}^{M} \frac{\bar{g}_{mnnm} - \bar{g}_{mmnn}}{4} \bigg( \frac{I - Z_{n}Z_{\omega} - Z_{m}Z_{\omega} + Z_{m}Z_{n} }{2} \bigg).
    \end{aligned}
\end{equation}

To get this equation in the form of Equation \eqref{eq:maxcut_cost}, we rewrite it as:

\begin{equation} \label{eq:JW4}
    \begin{aligned}  
        H_{MaxCut}^{\text{JW}}(R) &=  \sum_{n=1}^{M} \Bigg(-\bar{h}_{nn} + \bigg[\sum_{m=1}^{M} \frac{\bar{g}_{mnnm} - \bar{g}_{mmnn}}{4} \bigg] \Bigg
) \bigg(\frac{I - Z_{n}Z_{\omega}}{2}\bigg) + \sum_{\substack{m,n=1}}^{M}  \frac{\bar{g}_{mnnm} - \bar{g}_{mmnn}}{4} \bigg(\frac{ - Z_{m}Z_{\omega} + Z_{m}Z_{n}}{2} \bigg) \\
        &= \sum_{n=1}^{M} \Bigg(-\bar{h}_{nn} + \bigg[\sum_{m=1}^{M} \frac{\bar{g}_{mnnm} - \bar{g}_{mmnn}}{4} \bigg] \Bigg
) \bigg(\frac{I - Z_{n}Z_{\omega}}{2}\bigg) +\sum_{m=1}^{M} \Bigg(\bigg[\sum_{n=1}^{M} \frac{\bar{g}_{mnnm} - \bar{g}_{mmnn}}{4} \bigg] \Bigg) \bigg(\frac{ -Z_{m}Z_{\omega} }{2} \bigg) \\
&\; \; \;    + \sum_{\substack{m,n=1}}^{M}  \frac{\bar{g}_{mnnm} - \bar{g}_{mmnn}}{4} \bigg( \frac{Z_{m}Z_{n}}{2} \bigg) \\
        &= \sum_{n=1}^{M} \Bigg(-\bar{h}_{nn} + \bigg[\sum_{m=1}^{M} \frac{\bar{g}_{mnnm} - \bar{g}_{mmnn}}{4} \bigg] \Bigg
) \bigg(\frac{I - Z_{n}Z_{\omega}}{2}\bigg) +\underbrace{\sum_{n=1}^{M} \Bigg(\bigg[\sum_{m=1}^{M} \frac{\bar{g}_{nmmn} - \bar{g}_{nnmm}}{4} \bigg] \Bigg) \bigg(\frac{I-Z_{n}Z_{\omega} - I}{2} \bigg)}_{\text{relabeled $m \leftrightarrow n$ (as dummy index)}} \\
&\; \; \;    + \sum_{\substack{m,n=1}}^{M}  \frac{\bar{g}_{mnnm} - \bar{g}_{mmnn}}{4} \bigg( \frac{Z_{m}Z_{n}}{2} \bigg) \\
&= \sum_{n=1}^{M} \Bigg(-\bar{h}_{nn} + \bigg[\sum_{m=1}^{M} \frac{\bar{g}_{mnnm} - \bar{g}_{mmnn}}{4} + \frac{\bar{g}_{nmmn} - \bar{g}_{nnmm}}{4} \bigg] \Bigg) \bigg(\frac{I - Z_{n}Z_{\omega}}{2}\bigg)  \\
&\; \; \;    + \sum_{\substack{m,n=1}}^{M}  \frac{\bar{g}_{mnnm} - \bar{g}_{mmnn}}{4} \bigg( \frac{Z_{m}Z_{n}}{2} \bigg) -\sum_{n=1}^{M} \Bigg(\bigg[\sum_{m=1}^{M} \frac{\bar{g}_{nmmn} - \bar{g}_{nnmm}}{4} \bigg] \Bigg) \bigg(\frac{I}{2} \bigg)
    \end{aligned}
\end{equation}

Finally, we can simplify this as:

\begin{equation} \label{eq:maxcut_full2}
    \begin{aligned}  
        H_{MaxCut}^{\text{JW}}(R)  &= \sum_{n=1}^{M} \big[-\bar{h}_{nn} + G_{n} \big] \bigg(\frac{I - Z_{n}Z_{\omega}}{2}\bigg)  + \sum_{\substack{m,n=1}}^{M}  \frac{\bar{g}_{mnnm} - \bar{g}_{mmnn}}{4} \bigg(\frac{Z_{m}Z_{n}}{2}  \bigg) -  c\frac{I}{2}\\
        &= \sum_{n=1}^{M} \big[-\bar{h}_{nn} + G_{n} \big] \bigg(\frac{I - Z_{n}Z_{\omega}}{2}\bigg)  + \sum_{\substack{m,n=1}}^{M}  \underbrace{\frac{\bar{g}_{mmnn}- \bar{g}_{mnnm}}{4}}_{\times -1} \underbrace{\bigg(\frac{-Z_{m}Z_{n}}{2}}_{\times -1}  \bigg) -  c\frac{I}{2}   \\
        &= \sum_{n=1}^{M} \big[-\bar{h}_{nn} + G_{n} \big] \bigg(\frac{I - Z_{n}Z_{\omega}}{2}\bigg)  + \sum_{\substack{m,n=1}}^{M}  \frac{\bar{g}_{mmnn}- \bar{g}_{mnnm}}{4} \bigg(\frac{I-Z_{m}Z_{n}}{2}  - \frac{I}{2} \bigg) - c\frac{I}{2},
    \end{aligned}
\end{equation}
where:

\begin{equation} \label{eq:G_jw}
    \begin{aligned}  
        G_{n} &= \bigg[\sum_{m=1}^{M} \frac{\bar{g}_{mnnm} - \bar{g}_{mmnn}}{4} + \frac{\bar{g}_{nmmn} - \bar{g}_{nnmm}}{4} \bigg], \\
        c &= \sum_{n=1}^{M} \Bigg(\bigg[\sum_{m=1}^{M} \frac{\bar{g}_{nmmn} - \bar{g}_{nnmm}}{4} \bigg] \Bigg).
    \end{aligned}
\end{equation}
Overall, the MaxCut SCF problem is written as:

\begin{equation} \label{eq:maxcut_full}
    \begin{aligned}  
        H_{MaxCut}^{\text{JW}}(R)  
        &= \sum_{n=1}^{M} \big[-\bar{h}_{nn} + G_{n} \big] \bigg(\frac{I - Z_{n}Z_{\omega}}{2}\bigg)  + \sum_{\substack{m,n=1}}^{M}  \frac{\bar{g}_{mmnn}- \bar{g}_{mnnm}}{4} \bigg(\frac{I-Z_{m}Z_{n}}{2} \bigg),
    \end{aligned}
\end{equation}
where $\omega = 0$. We see that Equation \eqref{eq:maxcut_full} is in the same form as the MaxCut cost function. The coefficients in this Hamiltonian give the MaxCut graph edge weights.

We remark that in Equation \eqref{eq:maxcut_full}, $\big(\bar{g}_{mmnn} (R) - \bar{g}_{mnnm}(R) \big) \geq 0 \; \forall m,n$ and thus the number of positive edge weights scales as $M^{2}-M \approx \mathcal{O}(M^{2})$.  The number of negative edge weights will scale as $\mathcal{O}(M)$ due to the first term in this equation. This scaling is fortunate as many graph based algorithms require  positive edge weights only and thus it can be argued that to a good approximation the problem is representable by a positive graph. Note work in \cite{babbush2015chemical} (see equations $23$ and $24$) can be used to estimate the worst case scenario. 

The MaxCut graph problem defined in Equation \eqref{eq:maxcut_full}  represents the same problem as the diagonal fermionic Hamiltonian (equation  \ref{eq:HF_H2}), except a single extra ancilla/auxilary bit for the encoding was required. However, the derivation here contains no penalization terms that can be used to favor particular symmetries - for example the number of alpha and beta electrons and/or spin multiplicity and/or point group symmetries. If penalization terms are added to the QUSO/QUBO cost function then further negative edge weights can occur. However, an analysis will depend on the particular choice of symmetries and thus we omit a review. The steps outlined in this appendix should be easy to generalize for such problem-dependent situations.

Finally, as a reference we write down the penalty term for the number operator for the MaxCut problem: 
\begin{equation} \label{eq:penalty_number}
    \begin{aligned}  
        P &= \bigg(\sum_{j=1}^{M}\alpha_{j}^{\dagger} \alpha_{j}  - N\bigg)^{2} \\
        & \mapsto \\
        P_{\omega} &= \bigg(\sum_{j=1}^{M} \bigg[ \alpha_{j}^{\dagger} \alpha_{j} + \alpha_{\omega}^{\dagger} \alpha_{\omega} - 2 \alpha_{j}^{\dagger} \alpha_{j} \alpha_{\omega}^{\dagger} \alpha_{\omega} \bigg] - N\bigg)^{2}.
    \end{aligned}
\end{equation}
This will be two local under the Jordan-Wigner transform. This can easily be modified to go over only the alpha and beta indices to enforce the number of alpha and beta electrons respectively.

The diagonal part of the $S^{2}$ operator can be written as:
\begin{equation} \label{eq:S2_op}
    \begin{aligned}
S_{\text{D}}^{2} = \sum_{p=1}^{M/2} \frac{3}{4} &\bigg( a_{2p}^{\dagger} a_{2p} + a_{2p+1}^{\dagger} a_{2p+1} - a_{2p}^{\dagger} a_{2p} a_{2p+1}^{\dagger} a_{2p+1} - \\ 
&a_{2p+1}^{\dagger} a_{2p+1} a_{2p}^{\dagger} a_{2p}\bigg) + \\
 \sum_{p=1}^{M/2}\sum_{\substack{m=1 \\ m\neq p}}^{M/2} \frac{1}{4}   &\bigg( a_{2p}^{\dagger} a_{2p} a_{2m}^{\dagger} a_{2m} - a_{2p}^{\dagger} a_{2p} a_{2m+1}^{\dagger} a_{2m+1} -  \\
 &a_{2p+1}^{\dagger} a_{2p+1} a_{2m}^{\dagger} a_{2m} + a_{2p+1}^{\dagger} a_{2p+1} a_{2m+1}^{\dagger} a_{2m+1}\bigg)
    \end{aligned}
\end{equation}
where each sum goes over the spatial orbital indices (assumed even indices are spin-up and odd indices are spin-down). Under the Jordan-Wigner transform, this operator is quadratic. For $\langle S^{2}\rangle=0$, this operator times $\lambda$ can be added to the (diagonal) Hamiltonian to favor states with $S^{2}$ equal zero. $I_{\omega} \otimes S_{\text{D}}^{2}$ can be used in the MaxCut version of the problem.

\section{Approximation Algorithms}\label{sec:MaxCut}

The MaxCut problem can be stated as follows. Given a graph $G = (\mathcal{V}, \mathcal{E}) $
with edge weights $w_{ij} \in \mathbb{R}$,  maximize the following function:

\begin{equation} \label{eq:MaxCut}
\begin{aligned}
    \text{Maximize}  \bigg[ \frac{1}{2}\sum_{i<j}^{M} w_{ij} (1-z_{i} \cdot z_{j}) \bigg] \\
    \text{s.t. } z_{k} \in \{-1, 1 \}.
    \end{aligned}
\end{equation}
Here $\mathcal{V} \in \{1,2,\hdots, M\}$ is the set of graph vertices and $\mathcal{E}$ is a set where each element is a pair of distinct vertex indices that represents an edge between those vertices.

The goal of MaxCut is to partition $\mathcal{V}$ into two subsets $\mathcal{N}$ and $\mathcal{P}$ where all the vertices in the set $\mathcal{N}\subseteq \mathcal{V}$ have a $-1$ value and all the vertices in the set  $\mathcal{P} \subseteq \mathcal{V}$ have a $+1$ value  such that $\mathcal{P} \cap \mathcal{N} = \emptyset$ and  $\mathcal{P} \cup \mathcal{N} = \mathcal{V}$. This is represented by the vector $\vec{z}$, where the sign at index $i$ ($z_{i} \in \{ -1, +1\}$) determines if vertex $i$ is in set  $\mathcal{P}$ or $\mathcal{N}$. It should be clear that only adjacent vertices in opposite sets contribute to the MaxCut value.

MaxCut can be relaxed to the following problem:

\begin{equation} 
\begin{aligned}
    \text{Maximize}  &\bigg[ \frac{1}{2}\sum_{i<j}^{M} w_{ij} (1-v_{i} \cdot v_{j}) \bigg]  \\
    &\text{s.t. } v_{k} \in \mathbb{R}^{M}, \| v_{k} \| = 1.
    \end{aligned}
\end{equation}
This can be written as:

\begin{equation} \label{eq:SDP_X}
\begin{aligned}
    \text{Maximize}  &\bigg[ \frac{1}{2}\sum_{i<j}^{M} w_{ij} (1-X_{ij}) \bigg]\\
    \text{s.t. } &X \succcurlyeq 0 \\
    &[X]_{ii} = 1, \forall i = 1, \hdots, M.
    \end{aligned}
\end{equation}
Here $X$ is the Gram matrix of $\{ v_{1}, \hdots, v_{M} \}$, i.e. $X = V^{\dagger} V$ where each $v_{k}$ defines the $k$-th column of $V$. The constraint $\|v_{i}\|=1$ is expressed by requiring $[X]_{ii} = 1$. $X \succcurlyeq 0$ indicates $X$ is positive semidefinite (PSD), which is the case as $X$ equals $V^{\dagger} V$ by definition. Note the set of PSD matrices is a convex set.

We remark here that if the constraint $rank(X) =1$ is imposed on Equation \eqref{eq:SDP_X}, then the optimization becomes equivalent to the original problem. The approximation can be viewed as the relaxation of this non-convex rank constraint.

Once the SDP (Equation \eqref{eq:SDP_X}) is solved \footnote{This can be solved (up to $\epsilon$ accuracy) in time polynomial on the input size and $\log(\epsilon^{-1})$ \cite{goemans1994879}.}, an $X_{opt} = \Tilde{V}^{\dagger} \Tilde{V}$ is obtained, which defines a set of optimized vectors: $\{ \Tilde{v}_{k}\}_{k=1,\hdots,M}$ for the relaxed problem.  To generate a cut for the original problem, we sample a random hyperplane $\vec{r} \in \mathcal{R}^{M}$, where each component is drawn independently from a standard normal distribution: $r_{i} \sim \mathcal{N}(0,1)$. A cut $\vec{z}_{cut}$ is then defined by:

\begin{equation} \label{eq:soln}
\begin{aligned}
    \vec{z}_{cut} = sign\big[ \Tilde{V} \cdot \vec{r} \big],
    \end{aligned}
\end{equation}
where $sign$ rounds each element of the output vector to $+1$ and $-1$. In words, the vectors $\{ \Tilde{v}_{k}\}_{k=1,\hdots,M}$ (columns of $\Tilde{V}$) are separated into $+1$ and $-1$ sets depending on which side of the random hyperplane defined by $\vec{r}$ they lie on.

The Goemans-Williamson algorithm for the Max-Cut problem, which utilizes semidefinite programming (SDP) relaxation  (Equation \eqref{eq:SDP_X}) and hyperplane rounding (Equation \eqref{eq:soln}), achieves an approximation ratio of at least $\alpha_{G.W} = 0.878$ for graphs with positive edge weights \cite{goemans1994879, goemans1995improved}. This can be expressed as:




\begin{equation} \label{eq:summary}
\begin{aligned}
\underset{z_{i} \in \{-1, 1 \}}{\mathrm{{max}}}  \bigg[ \sum_{i<j}^{M} w_{ij} \bigg( \frac{1-z_{i}z_{j}}{2}  \bigg) \bigg] \geq \mathbb{E}[cut(H)] = \sum_{i<j}^{M} w_{ij} \bigg( Pr[ (i,j) 
\text{ in cut}]  \bigg) 
    &\geq  \alpha_{G.W}  \bigg[\sum_{i<j}^{M} w_{ij} \bigg(\frac{1-v_{i} \cdot {v}_{j}}{2} \bigg) \bigg] \\
    &\geq  \alpha_{G.W}  \bigg[\sum_{i<j}^{M} w_{ij} \bigg(\frac{1-\tilde{z}_{i} \cdot \tilde{z}_{j}}{2} \bigg) \bigg]
    \end{aligned}
\end{equation}
where $ \tilde{z}_{i} = sign[\Tilde{V} \cdot \vec{r}]_{i}$. The $\alpha_{G.W}=0.878$ ratio is derived from the expected value of the cut produced by the random hyperplane. Overall the expected value of the cut produced is at least $0.878$ times the optimal MaxCut value if $w_{ij}>0 \; \; \forall i,j$. However, for SCF problems the MaxCut graph can have negative edge weights. We can bound the expectation value in this scenario as follows. We write the following inequality:

\begin{equation} \label{eq:neg_weights}
\begin{aligned}
\underset{z_{i} \in \{-1, 1 \}}{\mathrm{{max}}}  \bigg[ \sum_{i<j}^{M} w_{ij} \bigg( \frac{1-z_{i} \cdot z_{j}}{2}  \bigg) \bigg] - W^{-}
    &\geq \alpha_{G.W}\Bigg(  \bigg[\sum_{i<j}^{M} w_{ij} \bigg(\frac{1-\tilde{z}_{i}\tilde{z}_{j}}{2} \bigg) \bigg] - W ^{-}   \Bigg),
    \end{aligned}
\end{equation}
where $W^{-}$ is a sum of all the negative edge weights:
\begin{equation} 
\begin{aligned}
W^{-} = \sum_{\substack{i,j \\ \forall w_{ij}<0}}^{M} w_{ij}.
    \end{aligned}
\end{equation}
We can add $W^{-}$ to both sides of Equation \eqref{eq:neg_weights} to obtain:

\begin{equation} \label{eq:neg_weights2}
\begin{aligned}
\underset{z_{i} \in \{-1, 1 \}}{\mathrm{{max}}}  \bigg[ \sum_{i<j}^{M} w_{ij} \bigg( \frac{1-z_{i}z_{j}}{2}  \bigg) \bigg] 
    &\geq \alpha_{G.W}  \bigg[ \sum_{i<j}^{M}w_{ij} \bigg(\frac{1-\tilde{z}_{i}\tilde{z}_{j}}{2} \bigg) \bigg] + (1- \alpha_{G.W}) W ^{-},
    \end{aligned}
\end{equation}
where $(1- \alpha_{G.W}) = 0.122$. We observe that the right-hand side of the inequality decreases in the presence of negative edge weights, indicating a slight deterioration in the approximation quality \cite{goemans1995improved, snir2012reconstructing, chatziafratis2021maximizing}.

\section{Atomic to Molecular Orbital Integrals}\label{sec:ao2mo_HF}

In this appendix we detail an efficient way to calculate all the molecular orbital integrals required to construct $H_{D}^{\text{ferm}}(\vec{\kappa})$ in the restricted orbital setting. This can easily be generalized to the open shell case if required.

To convert the two electron repulsion integrals (ERI) from the atomic orbital (AO) basis - denoted as $(\mu \nu | \lambda \sigma )$ - to the molecular orbital (MO) basis, denoted by $(pq|rs)$, the following tensor contraction is required:
\begin{equation}\label{eq:pqrs}
    (pq|rs) = g_{pqrs} =  \sum_{\mu \nu \lambda \sigma}^{N_{s}} C_{\mu p} C_{\nu q} (\mu \nu | \lambda \sigma ) C_{\lambda r} C_{\sigma s}.
\end{equation}
Note that these integrals are over $N_{s}$ spatial orbitals (usually this is half the number of spin orbitals: $N_{s} = M/2$). This four-tensor can be efficiently calculated by the $4$ contraction steps summarized in Table \ref{tab:ao2mo_int_full} \cite{taylor1987integral}.

However, not all these integrals are required to build the diagonal fermionic Hamiltonian. Only the following terms are required:

\begin{equation}\label{eq:pqqp}
    (pq|qp) = W_{pq}^{(pq|pq)}= \sum_{\mu \nu \lambda \sigma}^{N_{s}} C_{\mu p} C_{\nu q} (\mu \nu | \lambda \sigma ) C_{\lambda q} C_{\sigma p},
\end{equation}
and
\begin{equation}\label{eq:ppqq}
    (pp|qq) = W_{pq}^{(pp|qq)} = \sum_{\mu \nu \lambda \sigma}^{N_{s}} C_{\mu p} C_{\nu p} (\mu \nu | \lambda \sigma ) C_{\lambda q} C_{\sigma q}.
\end{equation}
These are both $(N_{s} \times N_{s})$ $2$D matrices  rather than $(N_{s} \times N_{s}\times N_{s}\times N_{s})$ four-tensors like Equation \eqref{eq:pqrs}. Overall, rather than transforming all the atomic orbital integral into the MO basis only the necessary terms have been transformed (providing both speed and memory improvements). The required contraction steps are summarized in Table \ref{tab:ao2mo_pqpq} and  Table \ref{tab:ao2mo_ppqq} respectively. 

\begin{table}[h!]
\begin{tabular}{ccc}
\hline
Scaling & Current       & Remaining           \\ \hline
$(N_{s})^5$       & $\mu \nu \lambda \sigma, \mu p \mapsto \nu \lambda \sigma p$ & $\nu q,\lambda r,\sigma s, \nu \lambda \sigma p \mapsto pqrs$ \\
$(N_{s})^5$       & $\nu \lambda \sigma p, \nu q \mapsto \lambda \sigma pq$  & $\lambda r, \sigma s, \lambda \sigma pq \mapsto pqrs$    \\
$(N_{s})^5$       & $\lambda \sigma pq, \lambda r \mapsto \sigma pqr$ & $\sigma s, \sigma pqr \mapsto pqrs$       \\
$(N_{s})^5$       & $\sigma pqr, \sigma s \mapsto pqrs$& $pqrs \mapsto pqrs$     \\ \hline
\end{tabular}
\caption{Optimal contraction path for $\mu p,\nu q,\mu \nu \lambda \sigma ,\lambda r,\sigma s \mapsto pqrs$, that gives all $(pq|rs)$ terms (Equation \eqref{eq:pqrs}). \label{tab:ao2mo_int_full}}
\end{table}

\begin{table}[h!]
\begin{tabular}{ccc}
\hline
Scaling & Current       & Remaining          \\ \hline
$(N_{s})^3$       & $\sigma s, \mu p \mapsto \sigma p \mu $    & $\nu q, \mu \nu \lambda \sigma, \lambda q, \sigma p \mu \mapsto pq $ \\
$(N_{s})^3$       & $\lambda q, \nu q \mapsto \lambda q \nu $    & $\mu \nu \lambda \sigma, \sigma p \mu, \lambda q \nu \mapsto pq $   \\
$(N_{s})^5$       & $\sigma p \mu, \mu \nu \lambda \sigma \mapsto p \nu \lambda$ & $\lambda q \nu, p \nu \lambda \mapsto pq $        \\
$(N_{s})^4$       & $p \nu \lambda, \lambda q\nu \mapsto pq$   & $pq \mapsto pq$             \\ \hline
\end{tabular}
\caption{Optimal contraction path for $\mu p,\nu q,\mu \nu \lambda \sigma ,\lambda p,\sigma q \mapsto pq$, that gives all $(pq|qp)$ terms (Equation \eqref{eq:pqqp}). \label{tab:ao2mo_pqpq}}
\end{table}

\begin{table}[h!]
\begin{tabular}{ccc}
\hline
Scaling & Current       & Remaining          \\ \hline
$(N_{s})^3$       & $\nu p, \mu p \mapsto \nu p \mu $    & $\mu \nu \lambda \sigma, \lambda q, \sigma q, \nu p \mu \mapsto pq $ \\
$(N_{s})^5$     & $\nu p \mu, \mu \nu \lambda \sigma \mapsto p \lambda \sigma $ & $\lambda q, \sigma q, p \lambda \sigma \mapsto pq$      \\
$(N_{s})^3$       & $\sigma q, \lambda q \mapsto \sigma q l$    & $p \lambda \sigma, \sigma q \lambda \mapsto pq $         \\
$(N_{s})^4$       & $\sigma q \lambda, p \lambda \sigma \mapsto pq $  & $pq \mapsto pq$            \\ \hline
\end{tabular}
\caption{Optimal contraction path for $\mu p,\nu p,\mu \nu \lambda \sigma ,\lambda q,\sigma q \mapsto pq$, that gives all $(pp|qq)$ terms (Equation \eqref{eq:ppqq}).\label{tab:ao2mo_ppqq}}
\end{table}
\clearpage

\section{SCF Numerical Data}
The following subsections provide the raw energies results for the \ce{OH-} and \ce{N2} studies respectively.

\subsection{\ce{OH-} data} \label{sec:OH_DATA}

Table \ref{tab:SCF_tab_OH} provides the numerical SCF results for the \ce{OH-} anion performed in conventional chemistry software. Table \ref{tab:alg1_data} and Table \ref{tab:alg2_data} provide the numerical results obtained from the SCF algorithms developed in this work.

\begin{table}[h!]
    \centering
\begin{tabular}{cccccc}
\hline
Psi4 & PySCF & Guassian & Guassian (QC) & ORCA & PySCF (SOSCF) \\
\hline
-74.216945 & -74.306277 & -74.224007 & -74.224007 & -74.835026 & -74.536030 \\
-74.256795 & -74.449520 & -74.584861 & -74.699716 & -74.955436 & -74.544027 \\
-74.408722 & -74.949772 & -74.813619 & -74.991018 & -75.012465 & -74.544115 \\
-75.023414 & -75.063266 & -75.062717 & -75.058227 & -75.178642 & -74.544115 \\
-75.072373 & -75.073001 & -75.070803 & -75.082348 & -74.917100 & -75.073002 \\
-75.073002 & -75.073002 & -75.072998 & -75.089361 & -75.095514 & -75.073002 \\
-75.073002 & -75.073002 & -75.073002 & -75.091783 & -75.121020 & -75.073002 \\
  & -75.073002 & -75.073002 & -75.093044 & -75.113305 & -75.087331 \\
 $-$ &  $-$ & -75.073002 & -75.093968 & -75.113306 & -75.107108 \\
 $-$ &  $-$ & -75.073002 & -75.094815 &  $-$ & -75.113279 \\
 $-$ &  $-$ & -75.073002 & -75.095650 &  $-$ & -75.113307 \\
 $-$ &  $-$ & -75.073002 & -75.096460 &  $-$ & -75.113307 \\
 $-$ &  $-$ &  $-$ & -75.097278 &  $-$ &  $-$ \\
 $-$ &  $-$ &  $-$ & -75.098068 &  $-$ &  $-$ \\
 $-$ &  $-$ &  $-$ & -75.098869 &  $-$ &  $-$ \\
 $-$ &  $-$ &  $-$ & -75.099622 &  $-$ &  $-$ \\
 $-$ &  $-$ &  $-$ & -75.100403 &  $-$ &  $-$ \\
 $-$ &  $-$ &  $-$ & -75.101127 &  $-$ &  $-$ \\
 $-$ &  $-$ &  $-$ & -75.101848 &  $-$ &  $-$ \\
 $-$ &  $-$ &  $-$ & -75.102520 &  $-$ &  $-$ \\
 $-$ &  $-$ &  $-$ & -75.103203 &  $-$ &  $-$ \\
 $-$ &  $-$ &  $-$ & ...        &  $-$ &  $-$        \\
 $-$ &  $-$ &  $-$ & -75.113307 &  $-$ &  $-$ \\
\hline
\end{tabular}
\caption{SCF data for \ce{OH-} described in the 6-31G basis set. Table \ref{tab:G_data} includes the full Guassian (QC) result. All energies are in Hartree (Ha).}
    \caption{}
    \label{tab:SCF_tab_OH}
\end{table}

\begin{table}[h!]
    \centering
\begin{tabular}{cccccccccc}
\hline
-74.224007 & -74.699716 & -74.991018 & -75.058227 & -75.082348 & -75.089361 & -75.091783 & -75.093044 & -75.093968 & -75.094815 \\
-75.095650 & -75.096460 & -75.097278 & -75.098068 & -75.098869 & -75.099622 & -75.100403 & -75.101127 & -75.101848 & -75.102520 \\
-75.103203 & -75.103831 & -75.104447 & -75.105016 & -75.105584 & -75.106103 & -75.106606 & -75.107066 & -75.107520 & -75.107931 \\
-75.108326 & -75.108684 & -75.109035 & -75.109351 & -75.109651 & -75.109923 & -75.110187 & -75.110425 & -75.110649 & -75.110851 \\
-75.111046 & -75.111221 & -75.111384 & -75.113250 & -75.113307 & -75.113307 & -75.113307 &  &  &  \\
\hline
\end{tabular}

    \caption{Guassian (QC) SCF data for \ce{OH-} described in the 6-31G basis set. All energies are in Hartree (Ha).}
    \label{tab:G_data}
\end{table}

\begin{table}[h!]
    \centering
\begin{tabular}{cccc}
\hline
CPLEX & Tabu  & Annealing  & MaxCut  \\
\hline
-75.062933 & -75.062933 & -75.062933 & -75.062933 \\
-75.113307 & -75.113307 & -75.113307 & -75.113307 \\
-75.113307 & -75.113307 & -75.113307 & -75.113307 \\
\hline
\end{tabular}
    \caption{MaxCut/QUBO-SCF data for \ce{OH-} described in the 6-31G basis set. Algorithm $1$ implementation - see Figure \ref{fig:alg_overview}. All energies are in Hartree (Ha).}
    \label{tab:alg1_data}
\end{table}

\begin{table}[h!]
    \centering
\begin{tabular}{cccc}
\hline
CPLEX  & Tabu  & Annealing  & MaxCut  \\
\hline
-74.275553 & -74.275553 & -74.275553 & -74.275553 \\
-75.089585 & -75.089585 & -75.089585 & -75.089585 \\
-75.103516 & -75.103516 & -75.103516 & -75.103516 \\
-75.111124 & -75.111124 & -75.111124 & -75.111124 \\
-75.112917 & -75.112917 & -75.112917 & -75.112917 \\
-75.113241 & -75.113241 & -75.113241 & -75.113241 \\
-75.113295 & -75.113295 & -75.113295 & -75.113295 \\
-75.113305 & -75.113305 & -75.113305 & -75.113305 \\
-75.113306 & -75.113306 & -75.113306 & -75.113306 \\
-75.113307 & -75.113307 & -75.113307 & -75.113307 \\
-75.113307 & -75.113307 & -75.113307 & -75.113307 \\
-75.113307 & -75.113307 & -75.113307 & -75.113307 \\
-75.113307 & -75.113307 & -75.113307 & -75.113307 \\
-75.113307 & -75.113307 & -75.113307 & -75.113307 \\
\hline
\end{tabular}
    \caption{MaxCut/QUBO-SCF data for \ce{OH-} described in the 6-31G basis set. Algorithm $2$ implementation - see Figure \ref{fig:alg_overview2}. All energies are in Hartree (Ha).}
    \label{tab:alg2_data}
\end{table}

\subsection{\ce{N2} data} \label{sec:N2_DATA}

Tables \ref{tab:SCF_N2_ccpvdz}, \ref{tab:SCF_N2_ccpvtz} and \ref{tab:SCF_N2_ccpvqz} provide the SCF results for the potential energy surface of molecular Nitgrogen. Tables \ref{tab:CISD_N2_ccpvdz},\ref{tab:CISD_N2_ccpvtz}and \ref{tab:CISD_N2_ccpvqz} provide the CISD numerical results.  Figure \ref{fig:N2_results-CISD} graphically shows these CISD results. The RHF and SO-SCF results are obtained from a calculation performed in PySCF. The internal stability was tracked to ensure the SO-SCF results converged to the true Hartree-Fock ground state.

\begin{table}[h!]
    \centering
\begin{tabular}{ccccccc}
\toprule
Bond length ({\AA}) & SO-RHF & RHF & MaxCut & Tabu & CPLEX & Annealing \\
\midrule
0.800000 & -108.407359 & -108.407359 & -108.407359 & -108.407359 & -108.407359 & -108.407359 \\
0.900000 & -108.782799 & -108.782799 & -108.782799 & -108.782799 & -108.782799 & -108.782799 \\
1.000000 & -108.929838 & -108.929838 & -108.929838 & -108.929838 & -108.929838 & -108.929838 \\
1.100000 & -108.953796 & -108.953796 & -108.953796 & -108.953796 & -108.953796 & -108.953796 \\
1.200000 & -108.914052 & -108.914052 & -108.914052 & -108.914052 & -108.914052 & -108.914052 \\
1.300000 & -108.843888 & -108.843888 & -108.843888 & -108.843888 & -108.843888 & -108.843888 \\
1.400000 & -108.761718 & -108.761718 & -108.761718 & -108.761718 & -108.761718 & -108.761718 \\
1.500000 & -108.679012 & -108.677514 & -108.677514 & -108.677514 & -108.677514 & -108.677514 \\
2.000000 & -108.468621 & -108.330583 & -108.442941 & -108.442941 & -108.330583 & -108.442941 \\
3.000000 & -108.310020 & -107.994079 & -108.308684 & -108.308684 & -108.308684 & -108.308684 \\
4.000000 & -108.240916 & -107.868563 & -108.240654 & -108.240654 & -108.226150 & -108.240654 \\
5.000000 & -108.216171 & -107.817283 & -108.216085 & -108.212826 & -108.212826 & -108.212826 \\
\bottomrule
\end{tabular}
    \caption{Potential energy surface data for the dissociation of \ce{N2} studied in the cc-pVDZ basis set performed at the SCF level of theory. All energies are in Hartree (Ha).}
    \label{tab:SCF_N2_ccpvdz}
\end{table}

\begin{table}[h!]
    \centering
\begin{tabular}{ccccccc}
\toprule
Bond length ({\AA}) & SO-RHF & RHF & MaxCut & Tabu & CPLEX & Annealing \\
\midrule
0.800000 & -108.498849 & -108.498849 & -108.498849 & -108.498849 & -108.498849 & -108.498849 \\
0.900000 & -108.839875 & -108.839875 & -108.839875 & -108.839875 & -108.839875 & -108.839875 \\
1.000000 & -108.968015 & -108.968015 & -108.968015 & -108.968015 & -108.968015 & -108.968015 \\
1.100000 & -108.983007 & -108.983007 & -108.983007 & -108.983007 & -108.983007 & -108.983007 \\
1.200000 & -108.939643 & -108.939643 & -108.939643 & -108.939643 & -108.939643 & -108.939643 \\
1.300000 & -108.868445 & -108.868445 & -108.868445 & -108.868445 & -108.868445 & -108.868445 \\
1.400000 & -108.786414 & -108.786414 & -108.786414 & -108.786414 & -108.786414 & -108.786414 \\
1.500000 & -108.704256 & -108.702748 & -108.702748 & -108.702748 & -108.702748 & -108.702748 \\
2.000000 & -108.492023 & -108.357519 & -108.465847 & -108.465847 & -108.357519 & -108.465847 \\
3.000000 & -108.336909 & -108.025870 & -108.335486 & -108.335486 & -108.025870 & -108.335486 \\
4.000000 & -108.272205 & -107.903173 & -108.254031 & -108.271927 & -107.903173 & -108.254031 \\
5.000000 & -108.245044 & -107.849205 & -108.240656 & -108.240217 & -107.849205 & -108.244958 \\
\bottomrule
\end{tabular}
    \caption{Potential energy surface data for the dissociation of \ce{N2} studied in the cc-pVTZ basis set performed at the SCF level of theory. All energies are in Hartree (Ha).}
    \label{tab:SCF_N2_ccpvtz}
\end{table}

\begin{table}[h!]
    \centering
\begin{tabular}{ccccccc}
\toprule
Bond length ({\AA}) & SO-RHF & RHF & MaxCut & Tabu & CPLEX & Annealing \\
\midrule
0.800000 & -108.515530 & -108.515530 & -108.515530 & -108.515530 & -108.515530 & -108.515530 \\
0.900000 & -108.851232 & -108.851232 & -108.851232 & -108.851232 & -108.851232 & -108.851232 \\
1.000000 & -108.976849 & -108.976850 & -108.976850 & -108.976850 & -108.976850 & -108.976850 \\
1.100000 & -108.990601 & -108.990601 & -108.990601 & -108.990601 & -108.990601 & -108.990601 \\
1.200000 & -108.946708 & -108.946708 & -108.946708 & -108.946708 & -108.946708 & -108.946708 \\
1.300000 & -108.875375 & -108.875375 & -108.875375 & -108.875375 & -108.875375 & -108.875375 \\
1.400000 & -108.793372 & -108.793372 & -108.793372 & -108.793372 & -108.793372 & -108.793372 \\
1.500000 & -108.711278 & -108.709785 & -108.709785 & -108.709785 & -108.709785 & -108.709785 \\
2.000000 & -108.498803 & -108.365684 & -108.365684 & -108.472576 & -108.365684 & -108.472576 \\
3.000000 & -108.344999 & -108.036377 & -108.343564 & -108.343564 & -108.036377 & -108.343564 \\
4.000000 & -108.282793 & -107.915404 & -108.282512 & -108.282512 & -107.915404 & -108.263479 \\
5.000000 & -108.255271 & -107.860868 & -108.249718 & -108.255181 & -107.860868 & -108.249235 \\
\bottomrule
\end{tabular}
    \caption{Potential energy surface data for the dissociation of \ce{N2} studied in the cc-pVQZ basis set performed at the SCF level of theory. All energies are in Hartree (Ha).}
    \label{tab:SCF_N2_ccpvqz}
\end{table}

\begin{table}[h!]
    \centering
\begin{tabular}{ccccccc}
\toprule
Bond length ({\AA}) & SO-RHF & RHF & MaxCut & Tabu & CPLEX & Annealing \\
\midrule
0.800000 & -108.648479 & -108.648479 & -108.648479 & -108.648479 & -108.648479 & -108.648479 \\
0.900000 & -109.040576 & -109.040576 & -109.040576 & -109.040576 & -109.040576 & -109.040576 \\
1.000000 & -109.204588 & -109.204588 & -109.204588 & -109.204588 & -109.204588 & -109.204588 \\
1.100000 & -109.246066 & -109.246065 & -109.246065 & -109.246065 & -109.246065 & -109.246065 \\
1.200000 & -109.224451 & -109.224451 & -109.224451 & -109.224451 & -109.224451 & -109.224451 \\
1.300000 & -109.172893 & -109.172893 & -109.172893 & -109.172893 & -109.172893 & -109.172893 \\
1.400000 & -109.109530 & -109.109529 & -109.109529 & -109.109530 & -109.109530 & -109.109530 \\
1.500000 & -109.034741 & -109.044012 & -109.044012 & -109.044012 & -109.044012 & -109.044012 \\
2.000000 & -108.765615 & -108.781719 & -108.717201 & -108.717201 & -108.781719 & -108.717201 \\
3.000000 & -108.642289 & -108.556716 & -108.640760 & -108.640760 & -108.640760 & -108.640760 \\
4.000000 & -108.619900 & -108.341826 & -108.619633 & -108.619633 & -108.616006 & -108.619633 \\
5.000000 & -108.615326 & -108.462398 & -108.615240 & -108.614685 & -108.614685 & -108.614685 \\
\bottomrule
\end{tabular}
    \caption{Potential energy surface data for the dissociation of \ce{N2} studied in the cc-pVDZ basis set performed at the CISD level of theory. All energies are in Hartree (Ha).}
    \label{tab:CISD_N2_ccpvdz}
\end{table}

\begin{table}[h!]
    \centering
\begin{tabular}{ccccccc}
\toprule
Bond length ({\AA}) & SO-RHF & RHF & MaxCut & Tabu & CPLEX & Annealing \\
\midrule
0.800000 & -108.837731 & -108.837731 & -108.837731 & -108.837731 & -108.837731 & -108.837731 \\
0.900000 & -109.187446 & -109.187445 & -109.187445 & -109.187445 & -109.187445 & -109.187445 \\
1.000000 & -109.325855 & -109.325855 & -109.325855 & -109.325855 & -109.325855 & -109.325855 \\
1.100000 & -109.352745 & -109.352745 & -109.352745 & -109.352745 & -109.352745 & -109.352745 \\
1.200000 & -109.322850 & -109.322851 & -109.322851 & -109.322851 & -109.322851 & -109.322851 \\
1.300000 & -109.266402 & -109.266402 & -109.266402 & -109.266402 & -109.266402 & -109.266402 \\
1.400000 & -109.199941 & -109.199940 & -109.199940 & -109.199940 & -109.199940 & -109.199940 \\
1.500000 & -109.121801 & -109.132172 & -109.132172 & -109.132172 & -109.132172 & -109.132172 \\
2.000000 & -108.851107 & -108.860368 & -108.809305 & -108.809305 & -108.860368 & -108.809305 \\
3.000000 & -108.723329 & -108.622654 & -108.721664 & -108.721664 & -108.376316 & -108.721664 \\
4.000000 & -108.695422 & -108.548108 & -108.690016 & -108.695134 & -108.548108 & -108.690016 \\
5.000000 & -108.688706 & -108.519979 & -108.688040 & -108.687628 & -108.519979 & -108.688618 \\
\bottomrule
\end{tabular}
    \caption{Potential energy surface data for the dissociation of \ce{N2} studied in the cc-pVTZ basis set performed at the CISD level of theory. All energies are in Hartree (Ha).}
    \label{tab:CISD_N2_ccpvtz}
\end{table}

\begin{table}[h!]
    \centering
\begin{tabular}{ccccccc}
\toprule
Bond length ({\AA}) & SO-RHF & RHF & MaxCut & Tabu & CPLEX & Annealing \\
\midrule
0.800000 & -108.909667 & -108.909668 & -108.909668 & -108.909668 & -108.909668 & -108.909668 \\
0.900000 & -109.251504 & -109.251504 & -109.251504 & -109.251504 & -109.251504 & -109.251504 \\
1.000000 & -109.385696 & -109.385696 & -109.385696 & -109.385696 & -109.385696 & -109.385696 \\
1.100000 & -109.410182 & -109.410182 & -109.410182 & -109.410182 & -109.410182 & -109.410182 \\
1.200000 & -109.378753 & -109.378754 & -109.378754 & -109.378754 & -109.378754 & -109.378754 \\
1.300000 & -109.321167 & -109.321168 & -109.321168 & -109.321168 & -109.321168 & -109.321168 \\
1.400000 & -109.253745 & -109.253743 & -109.253743 & -109.253743 & -109.253743 & -109.253743 \\
1.500000 & -109.174014 & -109.185092 & -109.185092 & -109.185092 & -109.185092 & -109.185092 \\
2.000000 & -108.903802 & -108.909556 & -108.909556 & -108.864203 & -108.909556 & -108.864203 \\
3.000000 & -108.772802 & -108.666549 & -108.771098 & -108.771098 & -108.666549 & -108.771098 \\
4.000000 & -108.741192 & -108.589876 & -108.740892 & -108.740892 & -108.589876 & -108.734925 \\
5.000000 & -108.733201 & -108.022607 & -108.732326 & -108.733110 & -108.560106 & -108.731826 \\
\bottomrule
\end{tabular}
    \caption{Potential energy surface data for the dissociation of \ce{N2} studied in the cc-pVQZ basis set performed at the CISD level of theory. All energies are in Hartree (Ha).}
    \label{tab:CISD_N2_ccpvqz}
\end{table}

\clearpage
\newpage
\section{SCF Algorithm overview} \label{sec:pseudo_algs}

Algorithms \ref{alg:cap1} and \ref{alg:cap2} provide the methodology for implementing the workflows depicted in Figure \ref{fig:algorithms}. Note, these are just the minimal approaches and other techniques, such as Direct Inversion of the Iterative Subspace (DIIS) \cite{pulay1980convergence}, can be used. Furthermore, penalization terms should be added each time $H_{D}$ is defined, as discussed in Appendix \ref{sec:mol_sym}. What terms are added will be problem dependent.

In Algorithm \ref{alg:cap2}, $J$, $K$ and $F$ are the Coulomb, exchange and Fock operators respectively. $S$ is the atomic orbital overlap matrix and  $H_{core}$ is the core-Hamiltonian \cite{szabo2012modern, helgaker2013molecular}. Note Algorithm \ref{alg:cap2} is written in terms of a restricted Hartree-Fock calculation, this can easily be extended to the restricted open-shell and unrestricted cases (by modifying how the Fock operator is constructed). How to do this is described in standard textbooks, such as  \cite{szabo2012modern, helgaker2013molecular}. 



\begin{algorithm}[H]
\caption{Implementation of Algorithm $1$ depicted in Figure \ref{fig:alg_overview}}\label{alg:cap1}
\begin{algorithmic}

\Require $\vec{\kappa}_{0},{h}_{pp},g_{ppqq},g_{pqqp}, \epsilon$ \Comment{$\epsilon$ is the convergence threshold}

\State $H_{D}(\vec{\kappa}) \gets$ Build Hamiltonian \Comment{See Equation \ref{eq:HF_H2} and \ref{eq:HF_H3}. Penalty terms can be added here to favour symmetries.}

\State $\ket{b_{min}} \gets  \text{argmin}_{\ket{b} \in  \mathcal{D}} \bigg[ \bra{b} H_{D}(\vec{\kappa}) \ket{b} \bigg]$ \Comment{\textbf{via: QUSO/QUBO/MaxCut/DQI/QAOA/QA}}

\Statex  

\State $E_{prev} \gets  \bra{b_{min}}H_{D}(\vec{\kappa}) \ket{b_{min}}$ 

\State $\text{converged} \gets  0$
\While{$\text{converged} \neq  1$}
\State $ \vec{\kappa} \gets  \text{argmin}_{\vec{\kappa}} \bigg[ \bra{b_{min}} H_{D}(\vec{\kappa}) \ket{b_{min}} \bigg]$  \Comment{via orbital gradient and Hessian \cite{helgaker2013molecular}}

\State $H_{D}(\vec{\kappa}) \gets$ Build Hamiltonian

\State $\ket{b_{min}} \gets  \text{argmin}_{\ket{b} \in  \mathcal{D}} \bigg[ \bra{b} H_{D}(\vec{\kappa}) \ket{b} \bigg]$ \Comment{\textbf{via: QUSO/QUBO/MaxCut/DQI/QAOA/QA}}

\Statex  

\State $E_{new} \gets  \bra{b_{min}}H_{D}(\vec{\kappa}) \ket{b_{min}}$ 

\If{$(E_{prev} - E_{new})<\epsilon$}
    \State $\text{converged} \gets  1$
\Else
    \State $E_{prev} \gets E_{new}$
\EndIf
\EndWhile

\Statex  

\State \textbf{Return:} $E_{new}$, $\vec{\kappa}$ and $\ket{b_{min}}$
\end{algorithmic}
\end{algorithm}

\begin{algorithm}[H]
\caption{Implementation of Algorithm $2$ depicted in Figure \ref{fig:alg_overview2}}\label{alg:cap2}
\begin{algorithmic}
\Require $H_{core},(\mu \nu | \lambda \sigma), \gamma^{initial}, S, \epsilon$ \Comment{$\epsilon$ is the convergence threshold}
\State $U, \Sigma, V^{\dagger} \gets \text{SVD}(S)$ \Comment{Perform singular value decomposition of $S$}
\State $A \gets U \Sigma^{-1/2}V^{\dagger}$

\Statex  

\State $J \gets \sum_{\lambda,\sigma} (\mu \nu | \lambda \sigma)\gamma_{\lambda \sigma}$
\State $K \gets  \sum_{\lambda,\sigma} (\mu \nu | \lambda \sigma) \gamma_{\lambda \sigma}$
\State $F \gets H_{core} + 2J - K$ \Comment{Build Fock matrix}
\Statex  
\State $F' \gets  A^{\dagger} F A$
\State $O,C' \gets \text{eigh}(F')$ \Comment{Get the eigenvalues  $O$ and eigenvectors $C'$ of the Hermitian matrix $F'$}

\State $C_{opt}  \gets  UC'$

\Statex  

\State $H_{D}(C^{opt}) \gets$ Build Hamiltonian \Comment{see Equation \eqref{eq:HF_HC}. Penalty terms can be added here to favour symmetries.}

\State  $\ket{b_{min}} \gets \text{argmin}_{\ket{b} \in  \mathcal{D}} \bigg[ \bra{b} H_{D}(C^{opt}) \ket{b} \bigg] $ \Comment{\textbf{via: QUSO/QUBO/MaxCut/DQI/QAOA/QA}}

\Statex  
\State $T_{pq} \gets \bra{b_{min}} (a_{2p}^{\dagger} a_{2q} + a_{2p+1}^{\dagger} a_{2q+1})   \ket{b_{min}}$  \Comment{spin-free $1$-RDM in MO basis}

\State $\gamma \gets C^{opt}T(C^{opt})^{\dagger}$ \Comment{transform  from molecular to atomic orbital basis ($T \mapsto \gamma$ )}

\Statex  

\State $J \gets \sum_{\lambda,\sigma} (\mu \nu | \lambda \sigma)\gamma_{\lambda \sigma}$
\State $K \gets  \sum_{\lambda,\sigma} (\mu \nu | \lambda \sigma) \gamma_{\lambda \sigma}$
\State $F \gets  H_{core} + 2J - K$

\State $E_{prev}   \gets  0.5 \sum_{\mu,\nu} \gamma_{\mu \nu} (H_{core} + F_{\mu \nu})$

\State $\text{converged} \gets  0$
\While{$\text{converged} \neq  1$}
    \State $F' \gets  A^{\dagger} F A$
    \State $O,C' \gets \text{eigh}(F')$ 
    
    \State $C_{opt}  \gets  UC'$

    \Statex  
    
    \State $H_{D}(C^{opt}) \gets$ Build Hamiltonian

    \State  $\ket{b_{min}} \gets \text{argmin}_{\ket{b} \in  \mathcal{D}} \bigg[ \bra{b} H_{D}(C^{opt}) \ket{b} \bigg] $ \Comment{\textbf{via: QUSO/QUBO/MaxCut/DQI/QAOA/QA}}

    \Statex  
    
    \State $T_{pq} \gets \bra{b_{min}} (a_{2p}^{\dagger} a_{2q} + a_{2p+1}^{\dagger} a_{2q+1})  \ket{b_{min}}$
    
    \State $\gamma \gets C^{opt}T(C^{opt})^{\dagger}$ 

    \Statex  
    
    \State $J \gets \sum_{\lambda,\sigma} (\mu \nu | \lambda \sigma)\gamma_{\lambda \sigma}$
    \State $K \gets  \sum_{\lambda,\sigma} (\mu \nu | \lambda \sigma) \gamma_{\lambda \sigma}$
    \State $F \gets  H_{core} + 2J - K$
    
    \State $E_{new}   \gets  0.5 \sum_{\mu,\nu} \gamma_{\mu \nu} (H_{core} + F_{\mu \nu})$

    \If{$(E_{prev} - E_{new})<\epsilon$}
        \State $\text{converged} \gets  1$
    \Else
        \State $E_{prev} \gets E_{new}$
    \EndIf
    
    \EndWhile
\Statex  
\State \textbf{Return:} $E_{new}$, $C^{opt}$ and $\ket{b_{min}}$
\end{algorithmic}
\end{algorithm}

\end{document}